\documentclass[twoside,english]{elsarticle}
\usepackage[T1]{fontenc}
\usepackage[latin9]{inputenc}
\usepackage{color}
\usepackage{array}
\usepackage{units}
\usepackage{multirow}
\usepackage{amsmath}
\usepackage{amsthm}
\usepackage{graphicx}
\usepackage{picins}
\PassOptionsToPackage{normalem}{ulem}
\usepackage{ulem}

\makeatletter

\providecommand{\tabularnewline}{\\}

\theoremstyle{plain}
\newtheorem{thm}{\protect\theoremname}
\theoremstyle{remark}
\newtheorem{rem}[thm]{\protect\remarkname}

\usepackage{lineno}

\@ifundefined{showcaptionsetup}{}{%
 \PassOptionsToPackage{caption=false}{subfig}}
\usepackage{subfig}
\makeatother

\usepackage{babel}
\providecommand{\remarkname}{Remark}
\providecommand{\theoremname}{Theorem}

\begin{document}

\begin{frontmatter}{}

\title{A Novel Communication Paradigm for High Capacity and Security via
Programmable Indoor Wireless Environments in Next Generation Wireless
Systems}

\author[forth]{Christos~Liaskos\corref{cor1}}

\ead{cliaskos@ics.forth.gr}

\author[gatech]{Shuai~Nie}

\ead{shuainie@ece.gatech.edu}

\author[forth]{Ageliki~Tsioliaridou}

\ead{atsiolia@ics.forth.gr}

\author[ucy]{Andreas~Pitsillides}

\ead{andreas.pitsillides@ucy.ac.cy}

\author[forth]{Sotiris~Ioannidis}

\ead{sotiris@ics.forth.gr}

\author[gatech,ucy]{Ian~Akyildiz}

\ead{ian@ece.gatech.edu}

\cortext[cor1]{Corresponding author}

\address[forth]{{\small{}Foundation for Research and Technology - Hellas (FORTH)}}

\address[gatech]{{\small{}Georgia Institute of Technology, School of Electrical and
Computer Engineering}}

\address[ucy]{{\small{}University of Cyprus, Computer Science Department}}
\begin{abstract}
Wireless communication environments comprise passive objects that
cause performance degradation and eavesdropping concerns due to anomalous
scattering. This paper proposes a new paradigm, where scattering becomes
software-defined and, subsequently, optimizable across wide frequency
ranges. Through the proposed programmable wireless environments, the
path loss, multi-path fading and interference effects can be controlled
and mitigated. Moreover, the eavesdropping can be prevented via novel
physical layer security capabilites. The core technology of this new
paradigm is the concept of metasurfaces, which are planar intelligent
structures whose effects on impinging electromagnetic waves are fully
defined by their micro-structure. Their control over impinging waves
has been demonstrated to span from $1$~GHz to $10$~THz. This paper
contributes the software-programmable wireless environment, consisting
of several HyperSurface tiles (programmable metasurfaces) controlled
by a central server. HyperSurfaces are a novel class of metasurfaces
whose structure and, hence, electromagnetic behavior can be altered
and controlled via a software interface. Multiple networked tiles
coat indoor objects, allowing fine-grained, customizable reflection,
absorption or polarization overall. A central server calculates and
deploys the optimal electromagnetic interaction per tile, to the benefit
of communicating devices. Realistic simulations using full 3D ray-tracing
demonstrate the groundbreaking performance and security potential
of the proposed approach in $2.4$~GHz and $60$~GHz frequencies.
\end{abstract}

\end{frontmatter}{}

\section{Introduction\label{sec:Intro}}

Recent years have witnessed a tremendous increase in the efficiency
of wireless communications. Multiple techniques have been developed
to tackle the stochastic nature of the wireless channel, in an effort
to fully adapt to its wide fluctuations. Indoor environments have
attracted special attention, since performance and security issues
accentuate due to the presence of multiple scatterers in a confined
space. In such cases, techniques such as MIMO, beamforming, adaptive
modulation and encoding have enabled wireless devices to rapidly adapt
to the time-variant, unpredictable channel state~\citep{pi2016millimeter}.
The present work opens an unexplored research path: making the wireless
environment fully controllable via software, enabling the optimization
of major propagation factors between wireless devices. Thus, effects
such as path loss, multi-path fading and interference become controllable,
allowing for novel capabilities in performance and physical-layer
security.

In order to understand the potential of exerting control over an environment,
we first need to define its composition and its natural behavior.
Indoor environments, which constitute the focus of the present work,
comprise two or more communicating devices\textendash such as laptops,
mobile phones, access points, base stations etc.\textendash and any
object found in a domestic or work space that can influence their
communication. At lower frequencies, walls, ceilings, floors, doors
and sizable furniture act as electromagnetic (EM) wave scatterers,
creating multiple paths between communicating end-points, especially
in non-line-of-sight (NLOS) areas. At higher frequencies, such as
millimeter wave (mm-wave) or terahertz ($THz$), which are expected
to play a major role in upcoming 5G communications~\citep{yilmaz2016millimetre},
even small objects act as substantial scatterers. Furthermore, ultra-small
wavelengths translate to considerable Doppler shift even at pedestrian
speed~\citep{yilmaz2016millimetre}. These factors, coupled with
the natural ambient dissipation of power due to free space losses,
lead to undermined NLOS performance at $2-5\,GHz$ and inability for
NLOS communications at $60\,GHz$ and beyond~\citep{yilmaz2016millimetre}.
Moreover, transmitted waves cannot be deterministically prevented
from reaching unintended recipients, causing interference and allowing
for eavesdropping.

\begin{figure}[t]
\begin{centering}
\includegraphics[width=1\columnwidth]{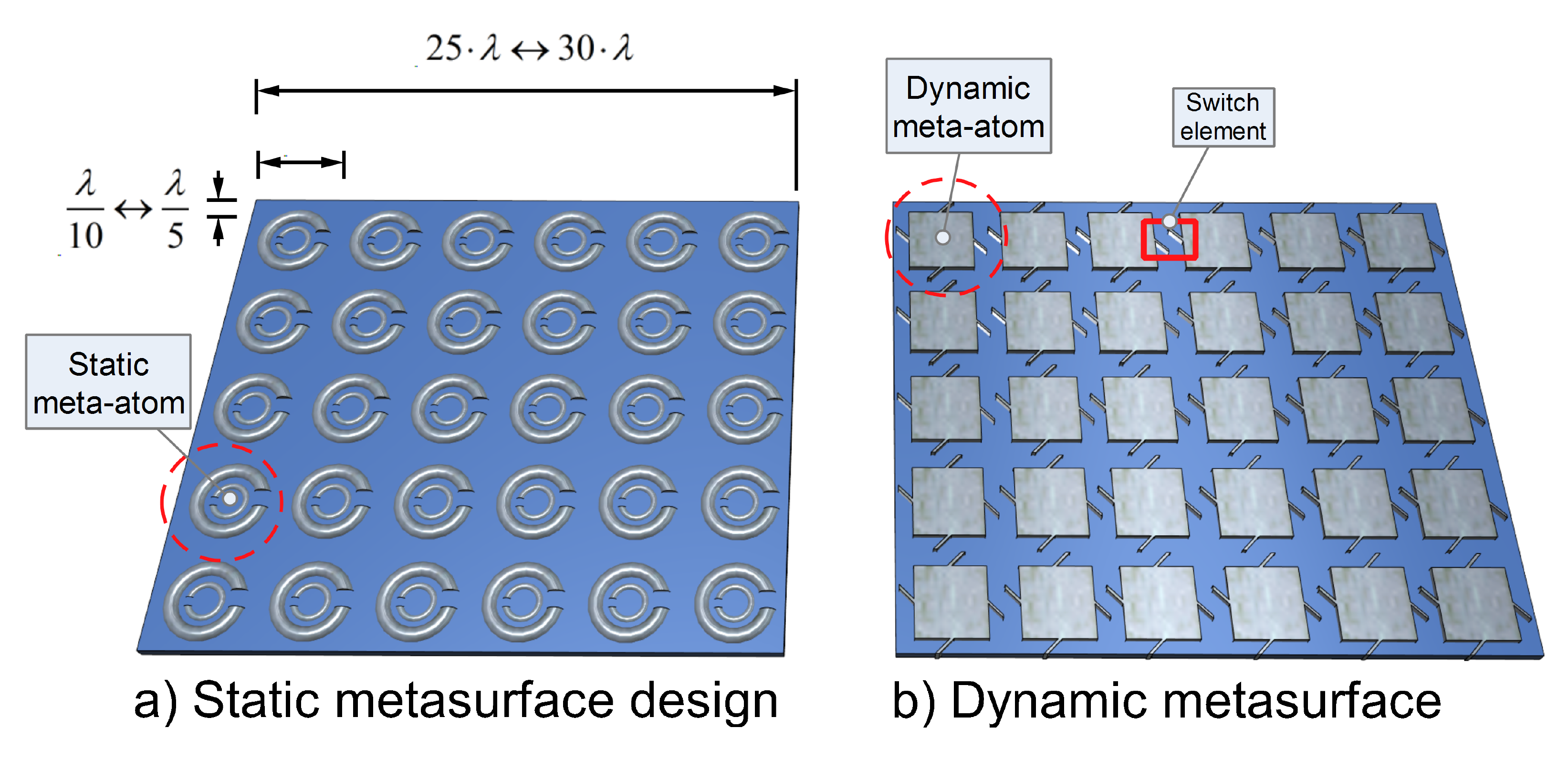}
\par\end{centering}
\caption{\label{fig:Mspatterns}Split ring resonators (left) constituted a
very common type of static metasurfaces, with fixed EM behavior. Novel
designs (right) incorporate switch elements (MEMS, CMOS or other)
to offer dynamically tunable EM behavior.}
\end{figure}

Existing proposals for physical-layer performance and security can
be classified as i) device-oriented, and ii) retransmitter-oriented.
Device-oriented methods include massive MIMO deployments in communicating
devices, to make constructive use of the multi-path phenomena~\citep{Aijaz.2017}.
Additionally, beamforming seeks to adaptively align the direction
of wireless transmissions in order to avoid redundant free space losses,
as well as to spatially contain interference and eavesdropping potential~\citep{kelif20163d,REFLECTARRAYS}.
Additional schemes include the on-the-fly selection of the modulation
and encoding scheme that offers the best bit error rate (BER) under
the current channel conditions~\citep{huang2017multi}. Retransmission-oriented
solutions advocate for the placement of amplifiers in key-positions
within the indoor environment. Retransmitters can be either passive
or active: Passive retransmitters are essentially conductive structures
akin to antenna plates~\citep{reflectInfocom.2017}. They passively
reflect energy from and towards fixed directions, without tunability.
Active retransmitters are powered electronic devices that amplify
and re-transmit received signals within a given frequency band. Essentially,
they attempt to combat power loss by diffusing more power within the
environment. In mm-wave frequencies and beyond, retransmitters must
be placed in line-of-sight (LOS) among each other, in an effort to
eliminate NLOS areas within a floor plan. Device-to-device networking
can also act as a retransmission solution for specific protocols and
a limited capacity of served users~\citep{chen2017promoting}. The
overviewed solutions have a common trait: They constitute device-side
approaches, which treat the environment as an uncontrollable factor
that does not participate into the communication process.
\begin{figure}[t]
\begin{centering}
\includegraphics[width=1\columnwidth]{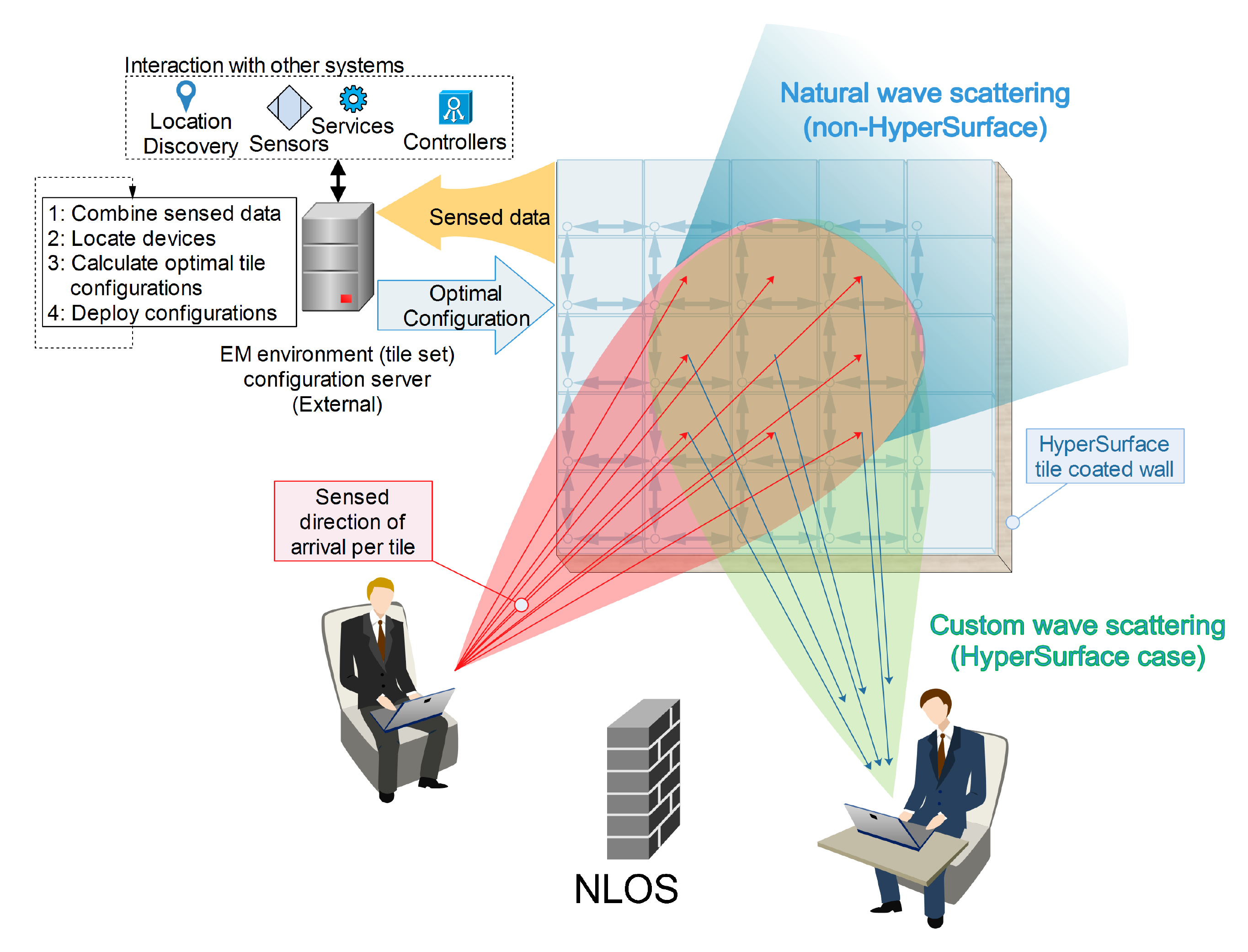}
\par\end{centering}
\caption{\label{fig:workflow}The proposed workflow involving HyperSurface
tile-coated environmental objects. The EM scattering is tailored to
the needs of the communication link under optimization. Unnatural
EM scattering, such as lens-like EM focus and negative reflection
angles can be employed to mitigate path loss and multi-path phenomena,
especially in challenging NLOS cases.}
\end{figure}
Metasurfaces are the core technology for introducing programmatically
controlled wireless environments~\citep{Zhu.2017,Minovich.2015,Lucyszyn.2010}.
They constitute the outcome of a research direction in Physics interested
in creating (rather than searching for) materials with required EM
properties. In their earlier iterations, they comprised a metallic
pattern, called \emph{meta-atom}, periodically repeated over a Silicon
substrate, as shown in Fig.~\ref{fig:Mspatterns}. The macroscopic
EM behavior of a metasurface is fully defined by the meta-atom form.
A certain pattern may fully absorb all impinging EM waves from a given
direction of arrival (DoA), while another may fully reflect a given
DoA towards another, at a negative reflection angle. Notably, metasurfaces
(and their 3D counterpart, the metamaterials) offer a superset of
EM behaviors with regard to regular materials. Lens functionality
(concentration of reflections towards a given point rather than ambient
dispersal) and negative refraction/reflection indices are some of
the exotic EM capabilities they can exhibit~\citep{Chen.2016}. Dynamic
meta-atom designs allow for dynamic metasurfaces, as shown in Fig.~\ref{fig:Mspatterns}.
Such designs include tunable factors, such as CMOS switches, microfluidic
switches or Micro Electro-Mechanical Switches (MEMS) that can alter
their state\textendash and the EM behavior of the metasurface\textendash via
an external bias~\citep{Chen.2016}. The bias is commonly electronic,
but thermal, light-based and mechanical approaches have been studied
as well~\citep{Chen.2016}. Thus, multi-functional metasurfaces,
that can switch from one EM behavior to another (e.g., from absorbing
to custom steering) are enabled. Finally, a very strong trait is that
there is no known limitation to the operating metasurface frequency,
which can be at the $mm$-wave and $THz$ bands~\citep{Lee.2012}.

The methodology proposed by the present study is to coat objects of
EM significance within an indoor environment with a novel class of
software-controlled metasurfaces. The study defines a unit of this
metasurface class, called \emph{HyperSurface tile}. A HyperSurface
tile is a planar, intelligent structure that incorporates networked
hardware control elements and adaptive meta-atom metasurfaces. Following
a well-defined programming interface, a tile can receive external
commands and set the states of its control elements to match the intended
EM behavior. The tiles, covering walls, doors, offices, etc., form
networks to facilitate the relaying of programmatic commands among
them. Moreover, tiles can have environmental sensing and reporting
capabilities, facilitating the discovery of communicating devices
within the environment. As shown in Fig.~\ref{fig:workflow}, a central
server can receive incoming tile reports, calculate the optimal configuration
per tile, and set the environment in the intended state by sending
the corresponding commands. Collaboration with existing systems (e.g.,
localization services and loud computing) constitutes a strong aspect
of the proposed approach, given that it enables the incorporation
of the EM behavior of materials in smart control loops.

The present study contributes the first model to describe programmable
wireless indoor environments, detailing their hardware, networking
and software components. The model includes the way for translating
EM metasurface functionalities to reusable software functions, bridging
physics and informatics. Moreover, the protocol specifications and
programming interfaces for interacting with tiles for communication
purposes are outlined. The practical procedure for deploying and configuring
programmable EM environments to mm-wave indoor communication is detailed.
The potential of programmable environments is evaluated via full 3D
ray tracing in $2.4$ and $60\,GHz$ cases, demonstrating their ground-breaking
potential in wireless performance and security.

The remainder of the text is organized as follows. Related studies
on physical-layer wireless performance and security are overviewed
in Section~\ref{sec:Related-Work}. Prerequisite knowledge on metasurfaces
is given in Section~\ref{sec:Backgrd}. The HyperSurface-based wireless
environment model is given in Section~\ref{sec:arch} and its configuration
is formulated in Section~\ref{sec:algorithm}. Applications to indoor
wireless setups are discussed in Section~\ref{sec:App}. Evaluation
via ray-tracing-based simulations is presented in Section~\ref{sec:Evaluation}.
Finally, the conclusion is  given in Section~\ref{sec:Conclusion}.

\section{Related Work\label{sec:Related-Work}}

Related to programmable wireless environments are the \emph{probabilistic
channel control} and \emph{physical layer security} concepts, surveyed
in the ensuing sub-sections.

\subsection{Probabilistic Channel Control}

The probabilistic channel control describes an approach to influence
the behavior of a communications channel not from its end-points (transmitter-receiver),
but rather from intermediate points. This has been exhibited with
the use of passive objects, phased array antenna panels and un-phased
antennas.

The position of passive objects naturally changes the propagation
of EM waves within a space. Spaces, such as floorplans, can even be
designed and built with EM wireless coverage considerations~\cite{wifihome}.
In existing spaces, metallic reflectors have been added as means of
naturally redirecting EM waves towards areas with poor coverage~\citep{reflectInfocom.2017}.
This approach does not offer adaptivity or precise control over EM
propagation. Moreover, the control type is limited to reflecting EM
waves to a natural direction. However, it is simple to deploy and
maintain in practice and cost-effective.

Phased array antennas have been used to actively and potentially adaptively
alter the probabilistic behavior of a channel. Array panels hung from
walls have been shown to influence considerably the communication
quality of wireless devices~\citep{kelif20163d,REFLECTARRAYS}. Phased
array antennas comprise several half- or quarter-wavelength antennas,
combined with hardware to control their relative phase. Altering the
relative phase of an antenna corresponds to a local change in the
reflective index of the array~\cite{yu2011light}. Thus, proper phase
configurations allow for anomalous wave steering and even absorbing.
However, the phase-based operation is coherent and deterministic only
at the far-field. For a square panel with size $D=0.5$~m and operating
frequency of $5$~GHz, the far field extends beyond $\nicefrac{2\left(\sqrt{2}\cdot D\right)^{2}}{\lambda}=16.67$~m.
For $60$~GHz the far field limit is at $200$~m. This constitutes
indoors applicability difficult, even for very small panels. Size-able
deployments can also be limited by the cost and power consumption
of the phase control hardware.

Un-phased antenna deployments have also been proposed as a cheaper
and simpler alternative. In this case, simple antennas are placed
over planar objects at relatively large distances to avoid coupling
effects. Control over the EM waves is exert only at the antenna positions,
while most of the surface of the planar object continues to interact
uncontrollably with EM waves. Thus, deterministic control is not attained,
even at the far-field. Instead, this approach attains a probabilistic
effect in the channel behavior, which can be quantified via measurements
after deployment has taken place~\cite{hotnetspapper}.

In differentiation, the present work proposes environments with software-defined,
\emph{deterministic wireless propagation}. This is attained by using
software-defined metasurfaces as the EM wave control agent~\citep{Liaskos.2015b}.
As detailed in Section~\ref{sec:Backgrd}, metasurfaces comprise
strongly coupled radiating elements sized at even less than $\nicefrac{\lambda}{10}$.
This high resolution of elements has been shown to allow for the micromanagement
of EM waves at the level of electric and magnetic field vectors, with
state-of-the-art spatial resolution and near unitary efficiency~~\citep{Banerjee.2011}.
This enables any kind of custom EM interaction, at any distance from
the surface, alleviating the deployment scalability concerns of phased
arrays. Moreover, their internal structure is simple: local refractive/reflective
index changes over the surface can be attained by simple ON/OFF switches.
This simplicity can enable cheap massive production, e.g., as printed
structures on films.

The supported features of the related approaches are summarized in
Table~\ref{table:comparisonQUA}.

\begin{table*}
\centering \caption{Comparison of EM wave control techniques.}
\label{table:comparisonQUA} \resizebox{\textwidth}{!}{%
\begin{tabular}{|c|c|c|c|c|c|}
\hline
 & \textbf{}%
\begin{tabular}{@{}c@{}}
\textbf{Far EM Field }\tabularnewline
\textbf{Control Type}\tabularnewline
\end{tabular}  & \textbf{}%
\begin{tabular}{@{}c@{}}
\textbf{Near EM Field}\tabularnewline
\textbf{Control Type}\tabularnewline
\end{tabular}  & \textbf{}%
\begin{tabular}{@{}c@{}}
\textbf{Spatial EM Control}\tabularnewline
\textbf{Granularity}\tabularnewline
\end{tabular}  & \textbf{}%
\begin{tabular}{@{}c@{}}
\textbf{Hardware }\tabularnewline
\textbf{Complexity}\tabularnewline
\end{tabular}  & \textbf{}%
\begin{tabular}{@{}c@{}}
\textbf{Deployment}\tabularnewline
\textbf{Scalability}\tabularnewline
\end{tabular} \tabularnewline
\hline
\textbf{}%
\begin{tabular}{@{}c@{}}
\textbf{Phased Array }\tabularnewline
\textbf{Antennas}\tabularnewline
\end{tabular}  & Deterministic & Probabilistic & Medium  & Highest & Lowest\tabularnewline
\hline
\textbf{}%
\begin{tabular}{@{}c@{}}
\textbf{Un-phased }\tabularnewline
\textbf{Antennas}\tabularnewline
\end{tabular}  & Probabilistic & Probabilistic & Low & Low & High \tabularnewline
\hline
\textbf{Passive Reflectors} & Probabilistic & Probabilistic & None & Lowest & Highest\tabularnewline
\hline
\textbf{HyperSurfaces} & Deterministic & Deterministic & Highest & Low & High\tabularnewline
\hline
\end{tabular} }
\end{table*}

\subsection{Physical Layer Security}

With the pervasive usage of smart wireless devices, security issues
have risen to become one of the most concerning aspects among end-users,
service providers, and policy makers worldwide. Information containing
personal credentials or with high security levels should be transmitted
and received in reliable channels against adversaries. To combat the
attacks of jamming and eavesdropping, traditional security techniques
are mostly deployed in the upper layers of the wireless networks,
for example, the Wi-Fi Protected Access (WPA) and WPA2 protocols in
IEEE 802.11 standards. In a modern cyber-physical system, security
methods are also being explored and implemented in the physical layer
where signal processing techniques and coding schemes are enhanced
for secrecy. In this Section, we briefly survey the state-of-the-art
in physical layer security techniques and make a brief comparison
and contrast with the HyperSurfaces, in terms of key enabling techniques
as well as computation complexity, as shown in Table~\ref{table:comparison}.

Major directions to achieve physical layer security (PLS) include
using highly directional antennas to nullify malicious attacks, forming
exclusion areas, assigning secret keys to legitimate users, and so
on. From the perspective of fundamental propagation channels, the
principle of a good secrecy can be achieved is when the eavesdroppers
do not have the knowledge of the frequencies where packages are transmitted,
or the eavesdroppers are in the same frequency channel but with much
higher noise which make the intercepted data impossible to decode~\citep{wyner1975wire}.
By understanding the attack patterns and corresponding combating strategies,
the secrecy capacity can be thus maximized in the wiretap channel~\citep{bloch2008wireless}.

\begin{table*}
\centering \caption{Comparison of Physical Layer Security Techniques.}
\label{table:comparison} \resizebox{\textwidth}{!}{%
\begin{tabular}{|c|c|c|c|c|c|}
\hline
 & \textbf{HyperSurface}  & \textbf{}%
\begin{tabular}{@{}c@{}}
\textbf{Millimeter-Wave }\tabularnewline
\textbf{Communications}\tabularnewline
\end{tabular}  & \textbf{}%
\begin{tabular}{@{}c@{}}
\textbf{Massive MIMO}\tabularnewline
\textbf{Communications}\tabularnewline
\end{tabular}  & \textbf{Channel Coding}  & \textbf{}%
\begin{tabular}{@{}c@{}}
\textbf{Heterogeneous}\tabularnewline
\textbf{Networks}\tabularnewline
\end{tabular} \tabularnewline
\hline
\textbf{Key Techniques}  & %
\begin{tabular}{@{}c@{}}
Deterministic Control \tabularnewline
of EM propagation\tabularnewline
\end{tabular} & %
\begin{tabular}{@{}c@{}}
Directional \tabularnewline
Beamforming\tabularnewline
\end{tabular} & Time Division Duplex  & %
\begin{tabular}{@{}c@{}}
LDPC, Polar Codes,\tabularnewline
Lattice Codes\tabularnewline
\end{tabular} & %
\begin{tabular}{@{}c@{}}
User Association Policies,\tabularnewline
Authentication and Authorization\tabularnewline
\end{tabular}\tabularnewline
\hline
\textbf{}%
\begin{tabular}{@{}c@{}}
\textbf{Computation }\tabularnewline
\textbf{Complexity}\tabularnewline
\end{tabular}  & Low  & High  & High  & High  & High \tabularnewline
\hline
\end{tabular}}
\end{table*}

\subsubsection*{Physical Layer Security on Millimeter-wave Communications}

Currently, several millimeter wave (mm-wave) frequency bands (30\textendash 300
GHz) are deployed for the next generation wireless communication systems~\citep{akyildiz20165g}.
With the advantage of more available spectrum resources, mm-wave systems
can achieve higher throughput compared to lower frequencies. However,
the limitation of higher path loss at mm-wave frequencies requires
the utilization of highly directional antennas or antenna arrays for
communication links to combat noise over a short distance. The characteristics
of high directivity and short-range communication are beneficial for
link security.

Recent studies have demonstrated that the security performance of
a mm-wave communication system relies on the antenna array patterns
and the density of eavesdroppers, and by introducing artificial noise
to the mm-wave system, the secrecy performance is significantly improved~\citep{wang2016physical,zhu2016physical,yang2015Safeguarding}.
However, challenges in fully utilizing mm-wave for the purpose of
PLS still remain. First, a comprehensive knowledge about mm-wave channel
is required for channel estimation, especially the peculiar effects
in blockage, atmospheric attenuation, and water vapor absorption.
Second, the computation efficiency in mm-wave beamforming needs to
be optimized under the scenario with multiple malicious attacks in
a small area.

\subsubsection*{Physical Layer Security on Massive MIMO Communications}

The massive MIMO communication system has the advantage of using very
large antenna arrays (with more than one hundred antenna elements)
at the transceivers to transmit or receive multiple streams of data
simultaneously. In terms of physical layer security, massive MIMO
system offers both benefits and drawbacks. On the positive side, the
channel condition is stable and easy to predict, which leads to a
reduced cost in channel estimation for users. Also, less complicated
signal processing burden is brought to both base stations (BSs) and
users. However, for eavesdroppers who are actively jamming the channels,
these advantages can also serve to their favors, which are the downside
of massive MIMO system that needs to be tackled~\citep{Kapetanovic2015physical}.

In order to achieve a desired secrecy level in massive MIMO communications,
one important step is to detect malicious activities. The active attacks
may forge themselves to act as legitimate users and hence intercept
the data from the base station. To solve this problem, several detection
strategies are discussed in~\citep{garnaev2014incorporating,wu2016secure,zhang2018pilot}.
For the eavesdroppers who are passively listening to the channel,
massive MIMO systems demonstrate good secrecy capacity by power allocation
scheme and artificial noise generation~\citep{zhu2014secure}.

\subsubsection*{Physical Layer Security on Heterogeneous Networks}

The architecture of heterogeneous networks (HetNets) allows for multiple
layers of cellular networks to operate with different coverage ranges,
transmit powers, radio access schemes, and so on. The extra degrees
of freedom in network configurations bring both opportunities and
challenges for physical layer security. On one hand, the network configurations
for high-power nodes and low-power nodes can be flexible and scalable
to account for different channel dynamics, including density of eavesdroppers,
mobility of authorized users, and channel fading, just to name a few~\citep{yang2015Safeguarding}.
On the other hand, the randomness in the HetNet brings the challenges
in authorized user association. For example, if a legitimate user
only selects the base station with the strongest power, it is also
easy for the eavesdroppers to intercept the information. Hence, a
tradeoff between link session connectivity and security requires the
design of user association schemes that also improve secrecy performance.

Additionally, with the burgeoning applications of wireless payment
and ad-hoc data exchange, device-to-device (D2D) communications in
very short distance among several users requires authentication and
authorization strategies to be carefully designed to avoid data leakage
to eavesdropping threats. Moreover, the D2D data relaying needs higher
security performance and reliable routing schemes to avoid data interception
and jamming~\citep{haus2017Security}. The computation complexity
for configuring an optimal route with high secrecy capacity will increase
as the number of relay nodes increases between two end devices.

\subsubsection*{Physical Layer Security Coding}

Coding schemes play a crucial role in improving a wireless system's
physical layer security. Existing codes for secrecy have been discussed
and surveyed extensively, which include low-density parity-check codes,
polar codes, lattice codes, among many others~\citep{harrison2013coding,poor2017wireless}.
The secrecy performance of different codes varies with different channel
conditions and types of eavesdropping behaviors. Therefore, challenges
still remain as in how to design codes that can maintain desirable
level of secrecy in more generalized channel and how coding schemes
can be synergistically combined with aforementioned techniques to
reach optimal results in future wireless systems.

\section{Prerequisites on Metasurfaces \label{sec:Backgrd}}

This section provides the necessary background knowledge on metasurfaces,
discussing dimensions and composition, operating principles and supported
functionalities. The following concise description targets a wireless
communications audience, given the topic of the present paper. A more
detailed introduction can be found in~\citep{Banerjee.2011}.

A metasurface is a planar, artificial structure which comprises a
repeated element, the meta-atom, over a substrate. In most usual compositions,
the meta-atom is conductive and the substrate is dielectric. Common
choices are copper over silicon, while silver and gold constitute
other exemplary conductors~\citep{Zhu.2017}. More exotic approaches
employ graphene, in order to interact with $THz$-modulated waves~\citep{Lee.2012}.
Metasurfaces are able to control EM waves impinging on them, in a
frequency span that depends on the overall dimensions. The size of
the meta-atom is much smaller than the intended interaction wavelength,~$\lambda$,
with $\nicefrac{\lambda}{10}-\nicefrac{\lambda}{5}$ constituting
common choices. The thickness of the metasurface is also smaller than
the interaction wavelength, ranging between $\nicefrac{\lambda}{10}\text{ and }\nicefrac{\lambda}{5}$
as a rule of a thumb. Metasurfaces usually comprise a dense population
of meta-atoms per area unit, which results into fine-grained control
over the EM interaction control. In general, a minimum size of approximately
$30\times30$ meta-atoms is required to yield an intended EM interaction~\citep{Chen.2016}.

Figure~\ref{fig:Mspatterns}-a illustrates a well-studied metasurface
design comprising split-ring resonators as the meta-atom pattern.
Such classic designs that rely on a static meta-atom, naturally yield
a static interaction with EM waves. The need for dynamic alteration
of the EM wave control type has given rise to dynamic metasurfaces,
illustrated in Fig.~\ref{fig:Mspatterns}-b. Dynamic meta-atoms incorporate
phase switching components, such as MEMS, CMOS transistors or microfluidic
switches, which can alter the structure of the meta-atom. Thus, dynamic
meta-atoms allow for time-variant EM interaction, while meta-atom
alterations may give rise to multi-frequency operation~\citep{Zhu.2017}.
Phase switching components can also be classified into state-preserving
or not. For instance, mechanical or microfluidic switches may retain
their state and require powering only for state transitions, while
semiconductor switches require power to maintain their state.

\begin{figure}[t]
\begin{centering}
\includegraphics[width=1\columnwidth]{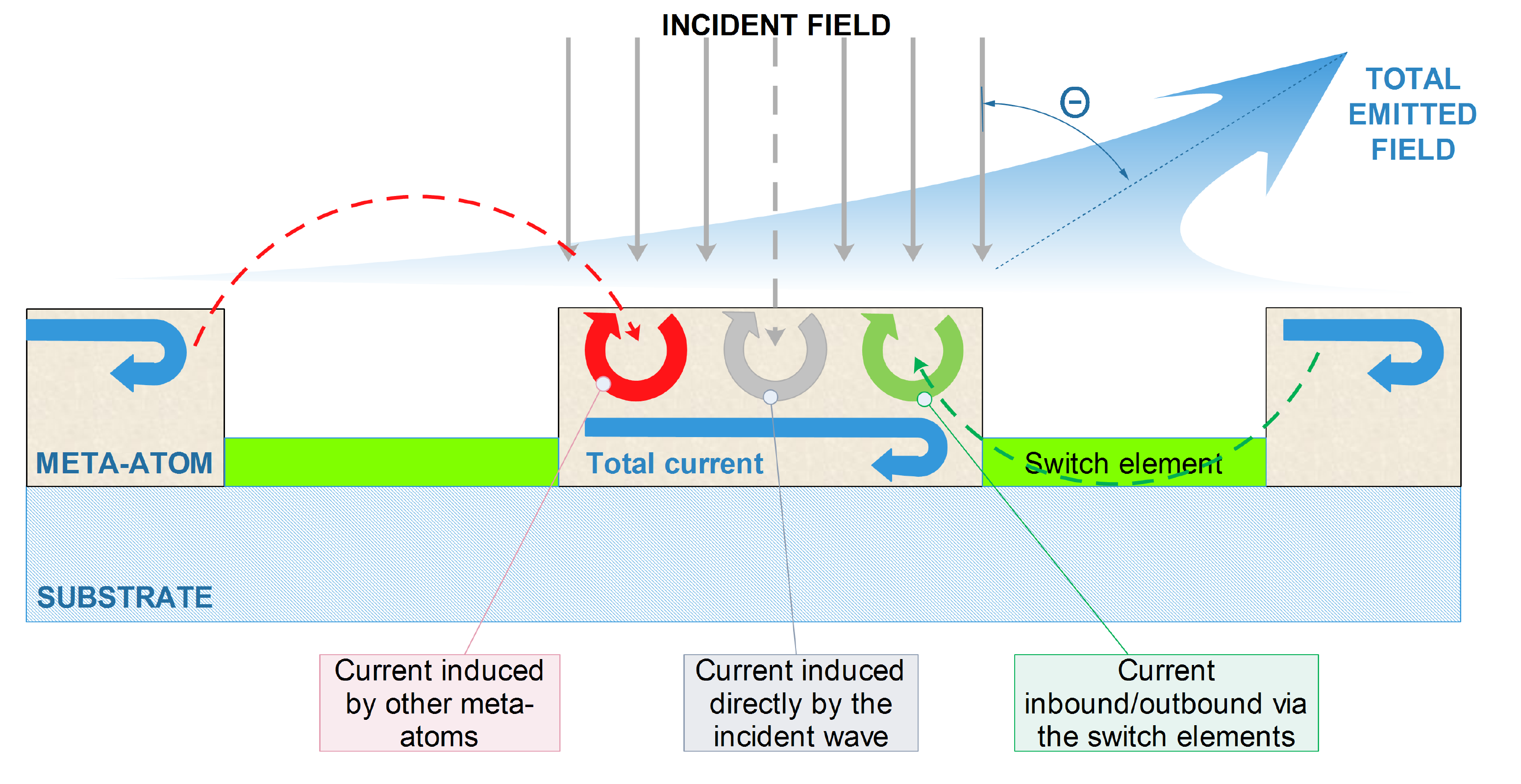}
\par\end{centering}
\caption{\label{fig:MSprinciple}The principle of metasurface functionality.
Incident waves create a well-defined EM response to the unit cells.
The cell response is crafted in such a way that the aggregate field
follows a metasurface-wide design objective, e.g., reflection towards
a custom angle $\Theta$.}
\end{figure}
The operating principle of metasurfaces is given in Fig.~\ref{fig:MSprinciple}.
The meta-atoms, and their interconnected switch elements in the dynamic
case, act as control factors over the surface currents flowing over
the metasurface. The total EM response of the metasurface is then
derived as the total emitted field by all surface currents, and can
take completely engineered forms, such as the unnatural reflection
angle shown in Fig.~\ref{fig:MSprinciple}. Engineering the total
surface current must account for all the currents over the surface.
These include: i) currents directly induced over the metasurface by
the incident wave, ii) currents induced in a meta-atom wirelessly
by other meta-atoms, and iii) currents flowing inwards or outwards
from a meta-atom via the switch elements. A qualitative description
of the dynamic metasurface operation can also be given: the meta-atoms
can be viewed as either input or output antennas, connected in custom
topologies via the switch elements. Impinging waves enter from the
input antennas, get routed according to the switch element states,
and exit via the output antennas, exemplary achieving customized reflection.

\subsection{State-of-the-art potential and manufacturing approaches\label{subsec:State-of-the-art-potential-and}}

Metasurfaces constitute the state of the art in EM control in terms
of capabilities and control granularity. A metasurface can support
a wide range of EM interactions, denoted as \emph{functions}. Common
function types include~\citep{Minovich.2015}:
\begin{itemize}
\item Redirection (refraction or reflection) of an impinging wave, with
a given direction of arrival, towards a completely custom direction.
Both the reflection and refraction functions can override the outgoing
directions predicted by Snell's law. Reflection and refraction functions
will jointly be referred to as wave \emph{steering}.
\item Beam splitting, i.e., steering a wave towards multiple custom directions
in parallel.
\item Wave absorbing, i.e., ensuring minimal reflected and/or refracted
power for impinging waves.
\item Wave polarizing, i.e., changing the oscillation orientation of the
wave's electric and magnetic field.
\item Wavefront focus, i.e., acting as lens to focus an EM wave to a given
point in the near or far field. Collimation (i.e., the reverse functionality)
can also be attained.
\item Phase control, i.e., altering the phase of the carrier wave.
\end{itemize}
Moreover, they can offer additional, advanced functions, such as anisotropic
response leading to hyperbolic dispersion relation, giant chirality,
arbitrary wave-front shaping and frequency selective filtering~\citep{Lucyszyn.2010}.
Apart from communications, these traits have been exploited in a variety
of applications, e.g., highly efficient energy harvesting photovoltaics,
and thermophotovoltaics, ultra-high resolution medical imaging, sensing,
quantum optics and military applications~\citep{Iwaszczuk.2012}.

The extended repertoire of EM function types, as well the exquisite
degree of granularity in EM behavior control, sets metasurfaces apart
from phased antennas and reflectarrays~\citep{PHASEDANTENNAS,REFLECTARRAYS},
which support coarser EM steering and absorbing at the far field,
e.g., for beamforming applications in wireless devices~\citep{kelif20163d}.
Notice that highly fine-grained EM control is required in mm-wave
setups, due to the extremely small wavelength~\citep{yilmaz2016millimetre}.

Regarding their manufacturing approaches, metasurfaces are commonly
produced as conventional printed circuit boards (PCBs)~\citep{Yang.2016}.
The PCB approach has the advantage of relying on a mature, commercially
accessible manufacturing technology. The PCB production cost is moderate
(indicatively, USD~$500$ per~$m^{2}$~\citep{PCBCART}). However,
the PCB technology is originally intended for integrated circuits
with far greater complexity than a metasurface. As described in the
context of Fig.~\ref{fig:Mspatterns}, a metasurface can be a very
simple structure, comprising a set of conductive patches, diodes and
conductive power/signal lines. Therefore, large area electronics (LAE)
can constitute better manufacturing approaches in terms of ultra low
production cost~\citep{LAEBOOK,LAEAPP}. LAE can be manufactured
using conductive ink-based printing methods on flexible and transparent
polymer films, and incorporate polymer/organic diodes~\citep{LAEPRINTED}.
Films with metasurface patterns and diodes printed on them can then
be placed upon common objects (e.g., glass, doors, walls, desks),
which may also act as the dielectric substrate for the metasurface.
It is theorized that printed electronics will reach the manufacturing
cost of regular paper printing~\citep{LAEcost}, which has an indicative
cost of USD~$1.66$~per~$m^{2}$~\citep{paper_print_cost}.

\section{The HyperSurface-based Programmable Wireless Environment Model\label{sec:arch}}

\begin{figure*}[t]
\begin{centering}
\includegraphics[width=1\textwidth]{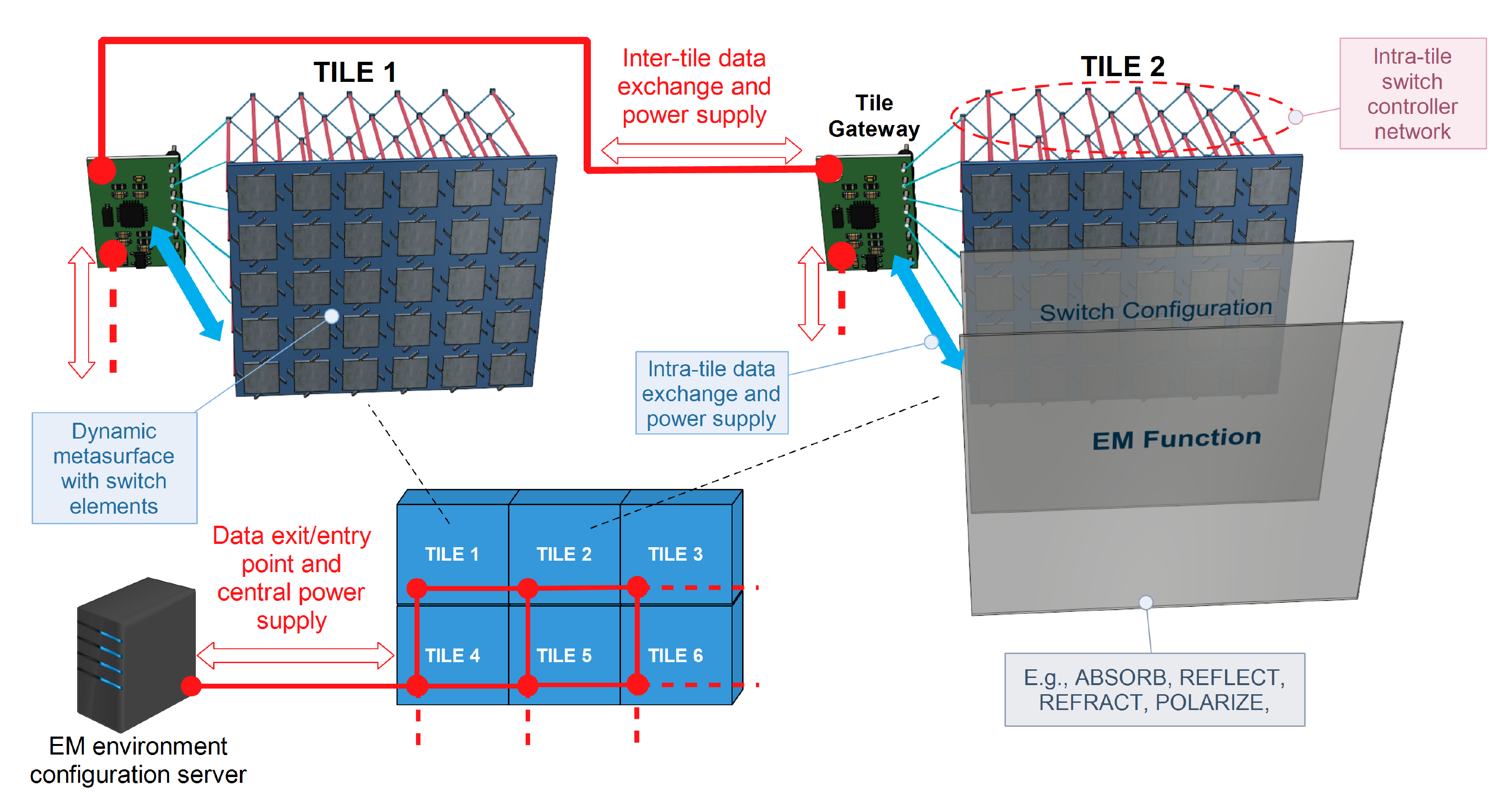}
\par\end{centering}
\caption{\label{fig:Architecture}Illustration of the HyperSurface tile architecture
and the tile-enabled wireless environment model. }
\end{figure*}
This section details the HyperSurface tile hardware components, the
tile inter-networking and the environment control software. A schematic
overview is given in Fig.~\ref{fig:Architecture} and is detailed
below.

\uline{The tile hardware}. The tile hardware consists of a dynamic
metasurface, a set of networked, miniaturized controllers that control
the switch elements of the metasurface, and a gateway that provides
inter-tile and external connectivity. The controller network has a
slave/master relation to the gateway. Via the gateway, the controller
network reports its current state and receives commands to alter the
state of the switch elements in a robust manner, making the metasurface
yield an overall required EM function.

A single controller is a miniaturized, addressable electronic device
that can monitor and modify the state of at least one metasurface
switch element. The controller design objectives are small size (to
avoid significant interference to the EM function of the metasurface),
low-cost (to support massive deployments in many tiles), high monitoring
and actuation speed (to sustain fast EM reconfigurability of the metasurface),
and the ability to create, receive and relay data packets (to enable
controller networking).

The avoidance of EM function disruption also refers to the wiring
required to connect the controllers to the switch elements and to
each other in a grid topology (cf. Fig.~\ref{fig:Architecture}).
Therefore, the total wiring should also be kept low. The grid-networked
controller approach is an option that balances wiring length and robustness
to node failures. Bus connectivity for the controllers would minimize
the required wiring, but would decrease the robustness against node
failures. On the other hand, a star connectivity would offer maximum
robustness but would also yield maximum wiring. Notice that future
technologies, such as nanonetworking, may enable wireless, computationally-powerful
nodes with autonomous, energy harvesting-based power supply~\citep{Akyildiz.2008}.
Thus, future tile designs may need no wiring or specific gateways.
The setup presented in this study prioritizes cost-effective realizability
with present manufacturing capabilities.

At a logical level, a controller is modeled as a finite-state automaton,
which reacts to incoming packets or switch element changes by transitioning
from one state to another~\citep{books2011automata}. A UML-standard
state diagram should capture three basic controller processes: the
data packet handling (including re-routing, consuming packets and
sending acknowledgments), the node reporting (reacting to an incoming
monitoring directive\textendash monitor request packet\textendash by
creating a new monitor data packet), and a fault detection process
(either self- or neighbor-failure). The latter is required for robust
data routing and for deducing the operational state of the tile as
a whole. Regarding the controller addressing, it can be either hardwired\textendash{}
due to the fixed grid topology\textendash or be set dynamically.

The tile gateway stands between the tile controller network and the
external world. It is incorporated to the tile fabric at a position
selected to yield minimal EM interaction concerns (e.g., at the back
of the tile). It provides mainstream protocol-compatible data exchange
with any other system. Internally, it is connected to at least one
controller, while more connections can be used for robust connectivity.
Moreover the gateway acts as a power supply bridge for the tile. Limited
size (e.g.,~\textasciitilde$cm$ ) and energy requirements are the
only significant constraints. Existing hardware, such as IoT platforms~\citep{Verikoukis.2017},
can be employed as tile gateways~\citep{Verikoukis.2017}. The tile
gateway may optionally have EM DoA sensing capabilities, to facilitate
the location discovery of wireless user devices in the environment.
\begin{rem}
It is noted that, in the simplest implementation, the gateway can
be directly wired to each metasurface switch, controlling them without
the intervention of a controller network. This control approach is
similar to LED arrays and monitors, and facilitates the printed film
manufacturing approaches described in Section~\ref{subsec:State-of-the-art-potential-and},
which can suffice for programmable wireless environments. The presence
of a controller network can enable advanced functions in the future,
such as self-maintained intelligent metasurfaces which can sense and
alter EM waves autonomously~\citep{Liaskos.2015b}.
\end{rem}
\uline{The tile inter-networking}. As tiles are placed over an
environmental object, such as a wall, they click together, connecting
data and power lines among the tile gateways (cf. Fig.~\ref{fig:Architecture}).
Thus, the tiles form a wired ad hoc network in a grid topology, where
existing IoT communication protocols can be readily employed. The
same protocol is used for connecting the tile network to any external
system. At least one tile\textendash denoted as exit/entry point\textendash has
its gateway connected to the environment configuration server, which
accumulates sensed data and diffuses EM actuation commands within
the tile network. More than one tile can be used as exit/entry points
at the same time, for the interest or robust and timely data delivery.

\uline{The environment control software}. The environment control
software is an application programming interface (API) that exists
at the configuration server. The API serves as a strong layer of abstraction,
hiding the internal complexity of the HyperSurfaces. It offers user-friendly
and general purpose access to metasurface functions, without requiring
knowledge of the underlying hardware and Physics. It provides software
descriptions of metasurface functions, allowing a programmer to customize,
deploy or retract them on-demand over tiles with appropriate callbacks.
These callbacks have the following general form:
\[
{\scriptstyle \texttt{outcome}\gets\texttt{callback(tile\_ID, action\_type, parameters)}}
\]
The $\texttt{tile\_ID}$ is the unique address of the intended tile
gateway in the inter-tile network (e.g., an IPv6). One EM function
per tile is considered here for simplicity. The $\texttt{action\_type}$
is an identifier denoting the intended function, such as $\texttt{STEER}$
or $\texttt{ABSORB}$, as described in Section~\ref{sec:Backgrd}.
Each action type is associated to a set of valid parameters. For instance,
$\texttt{STEER}$ commands require: i) an incident DoA, $\overrightarrow{I}$,
ii) an intended reflection direction, $\overrightarrow{O}$, and iii)
the applicable wavelength, $\lambda$, (if more than one are supported).
$\texttt{ABSORB}$ commands require no $\overrightarrow{O}$ parameter.
Notice that metasurface properties can be symmetric: i.e., a $\texttt{STEER}\left(\overrightarrow{I},\overrightarrow{O}\right)$
can also result into $\texttt{STEER}\left(\overrightarrow{O},\overrightarrow{I}\right)$~\citep{Holloway.2012}.

Once executed at the configuration server, a callback is translated
to an appropriate configuration of the switch elements that should
be deployed at the intended tile. The configuration is formatted as
a data packet that enters the tile network via an entry/exit point,
and is routed to the intended tile via the employed intra-tile routing
protocol. (An exemplary topology and routing strategy, considering
HyperSurface constraints, appears in~\citep{nocarc18}). The intended
tile gateway translates the directive according to the controller
network communication protocol specifications and diffuses it within
the tile. Upon success, it returns an acknowledgment to the configuration
server, or an error notification otherwise.

In the general case, the translation of an EM function to a tile switch
element configuration is accomplished via a lookup table, populated
during the tile design/manufacturing process as follows. Let $\sigma$
be a single tile configuration, defined as an array with elements
$s_{ij}$ describing the intended switch element state that is overlooked
by controller with address $i,j$ in the tile controller network.
(One-to-one controller-switch relation is assumed). In the MEMS case,
$s_{ij}$ takes binary values, $1$~or~$0$, denoting switch connection
or disconnection. Additionally, let $\Sigma$ be the set of all possible
configurations, i.e., $\sigma\in\Sigma$. Let an EM function of type
$\texttt{ABSORB}$ from DoA $\overrightarrow{I}$ be of interest.
Moreover, let $P_{\sigma}(\phi,\theta)$ be the power reflection pattern
of the tile (in spherical coordinates), when a wave with DoA $\overrightarrow{I}$
impinges upon it and a configuration $\sigma$ is active. Then, the
configuration $\sigma_{best}$ that best matches the intended function
$\texttt{ABSORB}\left(\overrightarrow{I}\right)$ is defined as:
\begin{equation}
\sigma_{best}\gets argmin_{\sigma\in\Sigma}\left\{ max_{\forall\phi,\theta}P_{\sigma}(\phi,\theta)\right\} \label{eq:fitness}
\end{equation}
Existing heuristic optimization processes can solve this optimization
problem for all functions of interest in an offline manner~\citep{haupt2007genetic},
using simulations or field measurements on prototypes. The configuration
lookup table is thus populated. Finally, we note that analytical results
for the EM function-configuration relation exist in the literature
for several metasurface designs~\citep{haupt2007genetic}. In such
cases, the analytical results can be employed directly, without the
need for lookup tables.

\section{Controlling Programmable Wireless Environments\label{sec:algorithm}}

In this Section we formulate the problem of optimally configuring
a programmable wireless environment to serve performance and security
objectives. The formulation is intended to facilitate the creation
of automatic environment configuration algorithms, as exemplary shown
later in Section~\ref{sec:Evaluation}.

\subsection{Wireless Performance Objectives}

In order to establish communication links between transmitters and
receivers, the HyperSurface tiles need to be adaptively selected and
optimally controlled to serve the desired receivers. Since in real-world
communication scenarios, multiple users can be present in the same
space, it is necessary to discuss the tile distribution and control
algorithms.

\textcolor{black}{We start from the case where there is one pair of
transmitter and receiver in the environment, as shown in Fig.~\ref{fig:WirCommExampleNegAngle}.
When the transmitter sends signals, multiple tiles can sense the transmitted
signals~\citep{nanocom.2017}. According to the location of the receiver
sensed by the location discovery system, those tiles will steer their
angles to establish reflection paths. Therefore, the signal received
at the receiver is a superposition of all signals reflected from various
tiles. In the particular case of Fig.~\ref{fig:WirCommExampleNegAngle},
among all tiles, only those that can sense the transmitted signals
need to respond to forwarding requests and tune their angles. On the
other hand, in a more complicated case where multiple users are present
in the same environment, the tile distribution needs to be optimized.
The signals to be received by different users should be orthogonal
to each other and are forwarded by different HyperSurface paths.}

\textcolor{black}{We assume the transmitted signal is QPSK modulated
with symbol $k(t)$, thus in general the received signal in time domain
can be expressed as
\begin{equation}
r(t)=k(t)\sum_{i=1}^{N}a_{i}e^{-j\theta_{i}}e^{j2\pi f_{c}\tau_{i}}+n(t),
\end{equation}
where $f_{c}$ is the central frequency, $a_{i}$, $\theta_{i}$,
and $\tau_{i}$ are the attenuation, phase, and delay caused by the
reflection paths along HyperSurface tiles of $i$-th path, and $n(t)$
is the AWGN noise in the channel. We assume there are in total $N$
paths found between the transmitter and receiver. The multipath effects
might cause distortion in overall received signal, therefore we need
to mitigate the destructive interference and harmonize the phases
by controlling the operation of HyperSurface tiles. Specifically,
we can formulate it as an optimization problem aimed at maximizing
the received power, $P_{r}^{(j)}$, and the number of tiles of the
HyperSurface, $M_{HS}^{(j)}$, for the $j^{\mathrm{th}}$ receiver
in the network with a total of $J$ users with $d_{j}$ distance,
as follows:
\begin{align}
\textit{Given:} & (x_{t},y_{t},z_{t}),(x_{r}^{(j)},y_{r}^{(j)},z_{r}^{(j)}),\\
 & P_{t}^{\mathrm{total}},M_{HS}^{\mathrm{total}}\\
\textit{Find: } & P_{t}^{(j)},M_{HS}^{(j)}\\
\textit{Objective: } & \max\sum d_{j}P_{r}^{(j)}\\
\textit{Subject to: }\nonumber \\
\text{Transmit power allocation: } & \sum P_{t}^{(j)}\leq P_{t}^{\mathrm{total}}\\
\text{HyperSurface tile allocation: } & \sum M_{HS}^{(j)}\leq M_{HS}^{\mathrm{total}}\\
\text{ for all }j\in J
\end{align}
In the above optimization problem, $(x_{t},y_{t},z_{t})$ and $(x_{r}^{(j)},y_{r}^{(j)},z_{r}^{(j)})$
denote the three-dimensional coordinates of the transmitter and the
$i^{\mathrm{th}}$ receiver, respectively. Based on the above optimization
problem, we can distribute tiles to corresponding users without causing
interference or signal distortion. }

\subsection{Physical-Layer Security Objectives\label{subsec:Physical-Layer-Security-Objectiv}}

We proceed to study two approaches for advanced physical layer security
in programmable wireless networks.

\textbf{Approach 1}. The first approach requires the deployment of
air-paths that cause zero or trivial interference to unintended users,
naturally blocking eavesdropping. The differentiation from the performance
objective is that the selection of air-paths prioritizes minimal interference
to unintended users, rather than maximal received power to the intended.
This can lead to improbable air-paths, e.g., long paths going around
crowded places, or paths confined above a given height within a floorplan.
We formulate this approach as follows:
\begin{align}
\textit{Given: } & (x_{t},y_{t},z_{t}),\\
 & (x_{r}^{(i)},y_{r}^{(i)},z_{r}^{(i)}),\\
 & (x_{r}^{(u)},y_{r}^{(u)},z_{r}^{(u)}),\\
 & P_{tx}^{\mathrm{tot}},N_{r}^{\mathrm{tot}},N_{t}^{\mathrm{tot}},\vec{n}_{t}^{initial}\\
\textit{Find: } & N_{t},\vec{n}_{t}^{(s)}\\
\textit{Objective: } & \min P_{r}^{(i,u)},\max P_{r}^{(i,i)}\\
\textit{Subject to: }\nonumber \\
\text{Transmit power allocation: } & \sum P_{t}^{(i)}\leq P_{tx}^{\mathrm{tot}}\\
\text{HyperSurface tiles allocation: } & N_{t}\leq N_{t}^{\mathrm{tot}}
\end{align}

In the above optimization problem, parameters $(x_{t},y_{t},z_{t})$,
$(x_{r}^{(i)},y_{r}^{(i)},z_{r}^{(i)})$, and $(x_{r}^{(u)},y_{r}^{(u)},z_{r}^{(u)})$
denote the three-dimensional coordinates of the transmitter, the intended
receiver, and unintended receivers, respectively. The total number
of receivers in the environment is denoted as $N_{r}^{\mathrm{tot}}$.
The initial condition of HyperSurface tiles are also known, which
is denoted as $\vec{n}_{t}^{initial}$. In order to minimize the received
power of unintended users trying to overhear $P_{r}^{(i,u)}$ while
maximizing the intended user's power $P_{r}^{(i,i)}$, we need to
find the number of tiles $N_{t}$ and adjust the tiles' orientation
to the correct angles, i.e., deploy a proper steering functionality.
In terms of formulation this can be seen as finding the normal vectors
of the selected tiles $\vec{n}_{t}^{(s)}$. The resources include
the transmit power that is upper-bounded by $P_{tx}^{\mathrm{tot}}$,
the number of HyperSurface tiles that are selected $N_{t}$ which
is bounded by the total number of tiles stays below $N_{t}^{\mathrm{tot}}$.

\textbf{Approach 2.} It is possible that a floorplan does not offer
air-paths that avoid other users completely. To address this case,
approach 2 seeks to ``scramble'' the reflected paths along propagation,
while still able to recover the original paths at the final bounces.
This can be done by altering the phases of the multipaths which belong
to the same cluster to achieve coherence, thus the signals' magnitude
can still be preserved. Upon the final bounce, the phase difference
should be minimized to zero to recover the signals. The objective
function is formulated as follows:
\begin{align}
\textit{Given: } & (x_{t},y_{t},z_{t}),\\
 & (x_{r}^{(i)},y_{r}^{(i)},z_{r}^{(i)}),\\
 & (x_{r}^{(u)},y_{r}^{(u)},z_{r}^{(u)}),\\
 & P_{tx}^{\mathrm{tot}},N_{r}^{\mathrm{tot}},N_{t}^{\mathrm{tot}},\vec{n}_{t}^{initial}\\
\textit{Find: } & N_{t},\vec{n}_{t}^{(s)}\\
\textit{Objective: } & \min P_{r}^{(i,u)},\\
 & \max P_{r}^{(i,i)},\\
 & \Delta\Phi_{n}^{(p,q)}=0\\
\textit{Subject to: }\nonumber \\
\text{Phase difference control: } & 0<\Delta\Phi_{j}^{(p,q)}\leq\pi/2,\\
 & \text{where}~1\leq p,q\leq N_{mpc}^{\mathrm{tot}},\\
 & \text{and}~1\leq j\leq n-1\\
\text{Transmit power allocation: } & \sum P_{t}^{(i)}\leq P_{tx}^{\mathrm{tot}}\\
\text{HyperSurface tiles allocation: } & N_{t}\leq N_{t}^{\mathrm{tot}}
\end{align}

In the above optimization problem, parameters $(x_{t},y_{t},z_{t})$,
$(x_{r}^{(i)},y_{r}^{(i)},z_{r}^{(i)})$, and $(x_{r}^{(u)},y_{r}^{(u)},z_{r}^{(u)})$
denote the three-dimensional coordinates of the transmitter, the intended
receiver, and unintended receivers, respectively. The total number
of receivers in the environment is denoted as $N_{r}^{\mathrm{tot}}$.
The initial condition of HyperSurface tiles are also known, which
is denoted as $\vec{n}_{t}^{initial}$. In order to minimize the received
power of unintended users trying to overhear $P_{r}^{(i,u)}$ while
maximizing the intended user's power $P_{r}^{(i,i)}$, we need
to find the number of tiles $N_{t}$ and adjust the tiles' orientation
to induce the phase changes $\Delta\Phi_{j}^{(p,q)}$ among multipaths
$p,q$ in the $j$-th reflection, which corresponds to find the normal
vectors of the selected tiles $\vec{n}_{t}^{(s)}$. The resources
include the phase difference controlled by the tiles, the transmit
power that is upper-bounded by $P_{tx}^{\mathrm{tot}}$, the number
of HyperSurface tiles that are selected $N_{t}$ which is bounded
by the total number of tiles stays below $N_{t}^{\mathrm{tot}}$.

\begin{figure}[t]
\begin{centering}
\includegraphics[width=1\columnwidth]{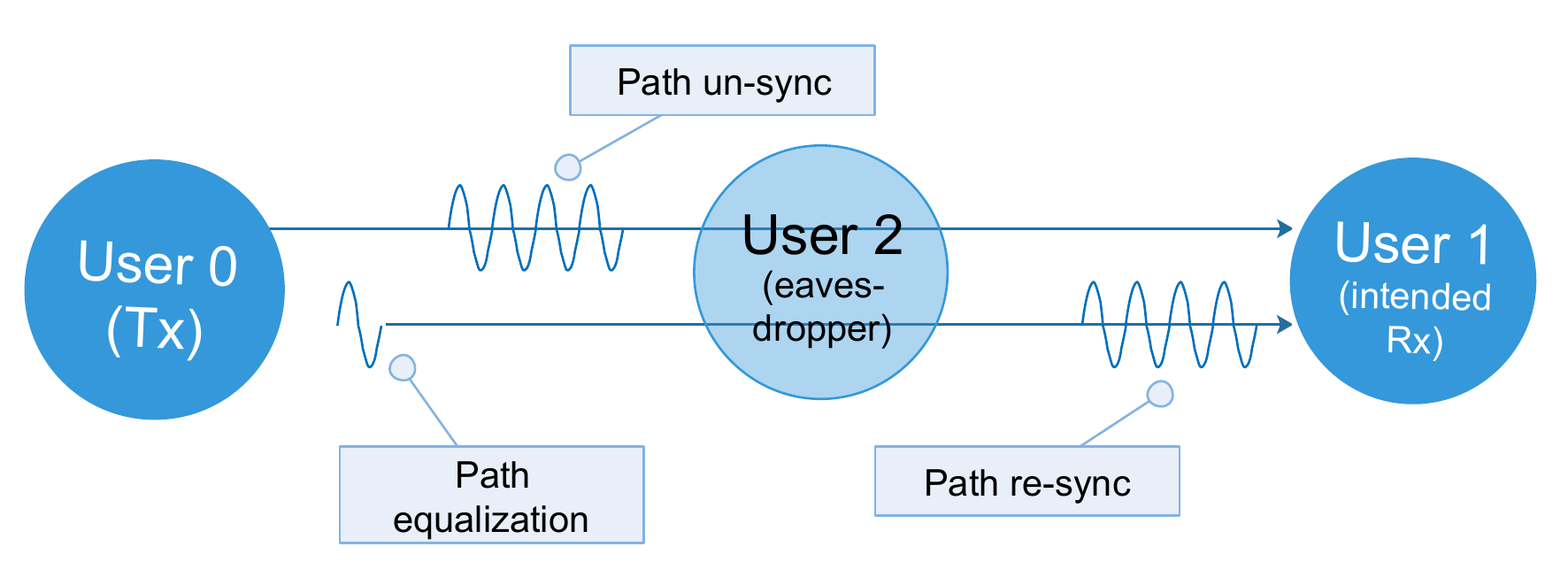}
\par\end{centering}
\caption{\label{fig:PhaseControlApproach}Illustration of the phase control-based
approach for eavesdropping mitigation. }
\end{figure}
The phase control-based approach is conceptually illustrated in Fig.~\ref{fig:PhaseControlApproach}.
We consider two users, a transmitter (user $0$) and the intended
receiver (user $1$), as well as an eavesdropped (user $2$) located
somewhere between two of the wave propagation paths. The paths are
assumed to carry approximately the same end-to-end power. At first,
a tile located over the path before the eavesdropper adds an equalization
phase to the fastest path, ensuring synchronicity at the user $1$.
Then, one of the paths is set to negative phase with regard to the
other, by using a tile located before user 2. This phase is canceled
out by using a tile between user $2$ and user $1$. In this manner,
the eavesdropper sees a total sum of zero received power, while the
reception is optimized at the intended user $1$.

\section{Applications to Mm-wave Indoor Setups\label{sec:App}}

\begin{figure}[t]
\begin{centering}
\includegraphics[width=1\columnwidth]{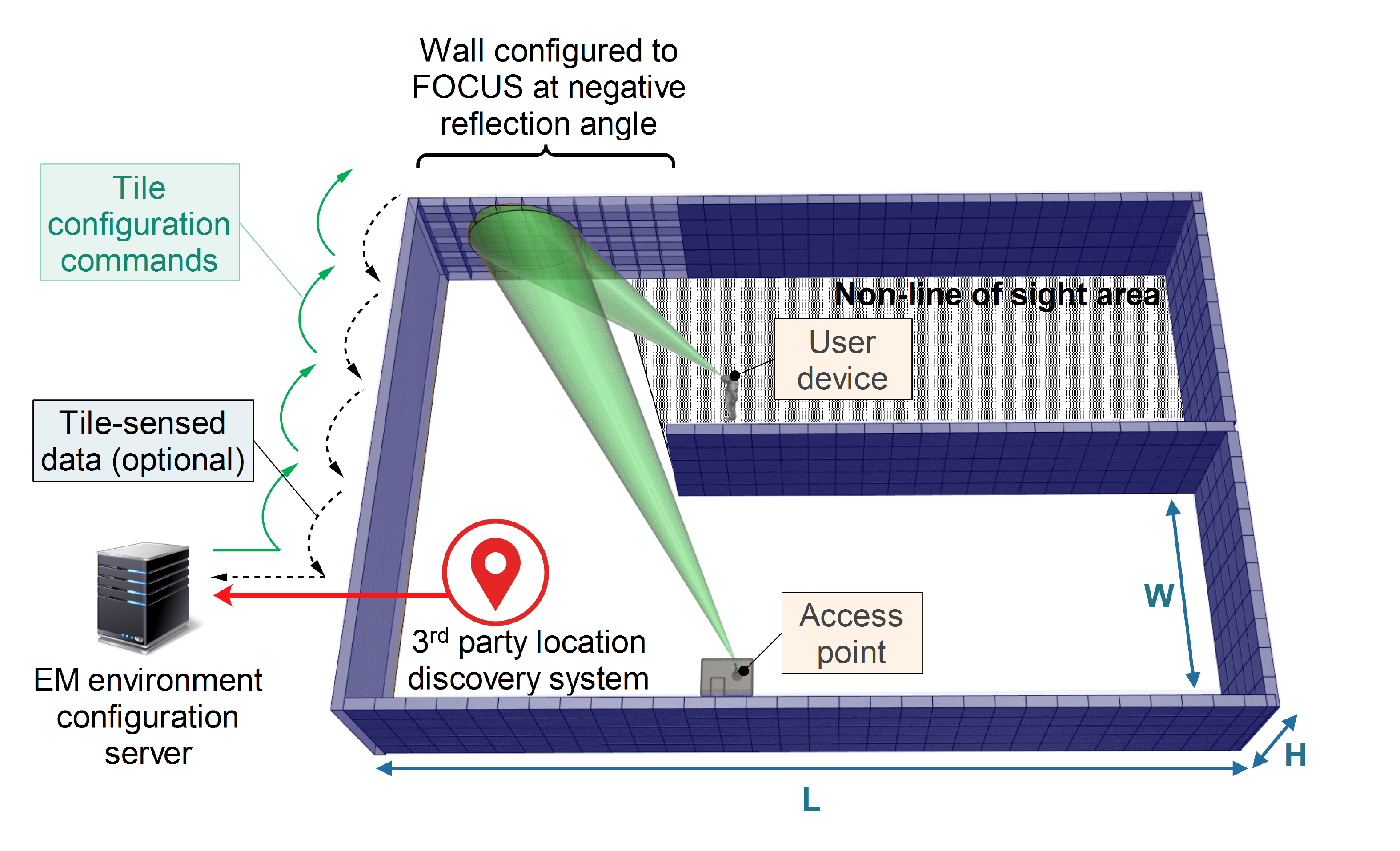}
\par\end{centering}
\caption{\label{fig:WirCommExampleNegAngle}Illustration of a customized wireless
indoor environment. $\texttt{STEER}$ functions are applied to several
tiles, to achieve a $\texttt{FOCUS}$ behavior of the corresponding
wall as a whole. }
\end{figure}
In mm-wave setups, major factors affect the signal attenuation: i)
the increased free space path loss (e.g., $\sim90\,dB$ at $10\,m$
for $60\,GHz$, instead of $60\,dB$ for $2.4\,GHz$), ii) acute multi-path
fading even in LOS cases, iii) strong Doppler shift even at pedestrian
speeds, and iv) optical-like propagation of EM waves, limiting connectivity
to LOS cases and exhibiting strong sensitivity to shadowing phenomena.
Attenuation due to molecular absorption may not play a significant
role in indoor cases\textendash depending on the composition of the
environment\textendash as it corresponds to $10^{-5}\,dB/m$ loss~\citep{pi2016millimeter}.

Given the mentioned mm-wave considerations, we proceed to present
mitigation measures offered by a HyperSurface-enabled environment.
We consider the setup of Fig.~\ref{fig:WirCommExampleNegAngle},
comprising a receiver (Rx)-transmitter (Tx) pair located in NLOS over
a known floorplan. The walls are coated with HyperSurface tiles. Furthermore,
we consider the existence of a location discovery service (e.g.,~\citep{localization60ghzCmaccuracy}),
which reports the location of the user device. At first, the Rx and
Tx may attempt high-power, omni-directional communication. The location
discovery service pinpoints the location of the user device and sends
it to the EM environment configuration server. (Without loss of generality,
the location of the Tx/access point can be considered known). Tiles
may sense their impinging power and report it to the server as well.
The server can use this information to increase the accuracy of the
discovered user device location. Subsequently, the following actions
take place:
\begin{itemize}
\item The tiles at the top-left part of Fig.~\ref{fig:WirCommExampleNegAngle}
are set to a symmetric ``negative focus'' setup as shown.
\item The Tx and the Rx are signaled to direct their antenna patterns to
the configured tiles using beamforming.
\end{itemize}
Unused tiles can be deactivated, reverting to regular, passive propagation.
Using this approach, the path loss can be even fully mitigated, since
the emitted energy is focused at the communicating end-points, rather
than scattering within the environment. This can also be of benefit
to the user device's battery lifetime, given that the redundantly
emitted power is minimized. Concerning multi-path fading, the fine-grained
EM control over the wave propagation can have as an objective the
\emph{crafting} of a power delay profile that mitigates the phenomenon,
e.g., by ensuring a path with significantly more power than any other,
or one that best matches the MIMO capabilities of the devices. Additionally,
the focal point of the EM wave reflected by the tiled wall towards
the use device can be altered in real-time, to match the velocity
of the mobile user. Mobile trajectory predictions can be employed
to facilitate this course of action. This provides a potential mitigation
approach for Doppler phenomena.

The environment optimization for multiple user pairs, or sub-spaces
within the environment, may be of increased practical interest. Returning
to the setup of Fig.~\ref{fig:WirCommExampleNegAngle}, the configuration
server can, e.g., set the tiles to preemptively minimize the delay
spread within the whole NLOS area, while ensuring a minimum level
of received power within it. In the sub-space optimization case, the
best matching tile configurations can be calculated offline and be
deployed upon request. This approach is evaluated in Section~\ref{sec:Evaluation}.

Finally, it is noted that the programmable environment extends the
communication distance of devices, without requiring extra dissipation
of energy within the environment (e.g. by placing additional access
points). This can constitute a considerable advantage for mm-wave
communications, which are known to be absorbable by living tissue.
Moreover, assuming tiles with state-preserving switch elements, the
energy footprint of the programmable environment can be extremely
low, especially in static or mildly changing user positions.

\section{Evaluation in $60\,GHz$ and $2.4\,GHz$ setups\label{sec:Evaluation}}

In this Section we evaluate the performance and security prospects
of programmable wireless environments. The evaluation employs full
3D, ray-tracing-based simulations. Different approaches for optimizing
the environment are demonstrated per each case.

\subsection{Performance objectives}

We proceed to evaluate the HyperSurface potential in mitigating the
path loss and multi\textendash path fading effects. The indoor 3D
space of Fig.~\ref{fig:WirCommExampleNegAngle} is ported to a full-3D
ray-tracing engine~\citep{ActixLtd.2010}, customized to take into
account HyperSurface tile functions. The evaluation focuses on finding
tile configurations that optimally mitigate the path loss and multi\textendash path
fading for $12$ users within the NLOS area. We study the case of
$60\,GHz$, which is of increased interest to upcoming 5G communications,
as well as the $2.4\,GHz$ case due to its wide applicability, e.g.,
to WiFi setups~\citep{yilmaz2016millimetre}.

Concerning the simulation parameters, the space has a height of $H=3\,m$,
corridor length (distance between opposite wall faces) $L=15\,m$,
corridor width $W=4.5\,m$, a middle wall length of $12\,m$, and
$0.5\,m$ wall thickness. Two stacked walls exist in the middle. The
floor and ceiling are treated as plain, planar surfaces composed of
concrete, without HyperSurface functionality. All walls are coated
with HyperSurface tiles, which are square-sized with dimensions $1\times1\,m^{2}$.
Thus, the 3D space comprises a total of $222$ tiles.

The dynamic metasurface pattern of Fig.~\ref{fig:Mspatterns} is
considered using state-preserving switches (e.g., microfluidic). Appropriate
dimensions are assumed, for $60\,GHz$ and $2.4\,GHz$ respectively,
as explained in the context of Fig.~\ref{fig:Mspatterns}. This pattern
design has been extensively studied in literature, offering a wide
range of steering and absorbing capabilities, even with switch elements
only at the horizontal direction~\citep[p. 235]{haupt2007genetic}.
Although beyond of the present scope, it is noted that this metasurface
design also exhibits tunable EM interaction frequency, yielding a
particularly extended repertoire of supported tunability parameters.
The considered tile functions account for EM wave steering and absorption
from various DoAs. Specifically, we allow for any DoA and reflection
direction resulting from the combination of $\left\{ -30^{o},\,-15{}^{o},\,0^{o},\,15^{o},\,30^{o}\right\} $
in azimuth and $\left\{ -30^{o},\,-15{}^{o},\,0^{o},\,15^{o},\,30^{o}\right\} $
in elevation planes, using the tile center as the origin. Notice that
the considered angles have been shown to be commonly attainable by
metasurfaces~\citep{Yazdi.2017}. However, carefully designed, static
metasurfaces have achieved nearly full angle coverage, i.e., almost
$\left(-90^{o},\,90^{o}\right)$ in azimuth and elevation, which is
indicative of their potential~\citep{Albooyeh.2014}. The reflection
coefficient is set to $100\%$ for each steering function~\citep[p. 235]{haupt2007genetic}.
Additionally, we consider an EM absorbing tile function which reduces
the power of impinging waves (given DoA) by $35\,dB$~\citep[p. 235]{haupt2007genetic},
scattering the remaining wave power towards the Snell's law-derived
reflection direction. Thus, a tile supports $26$ different function
configurations in total.
\begin{figure*}[t]
\begin{centering}
\includegraphics[viewport=50bp 240bp 550bp 650bp,clip,width=0.49\textwidth]{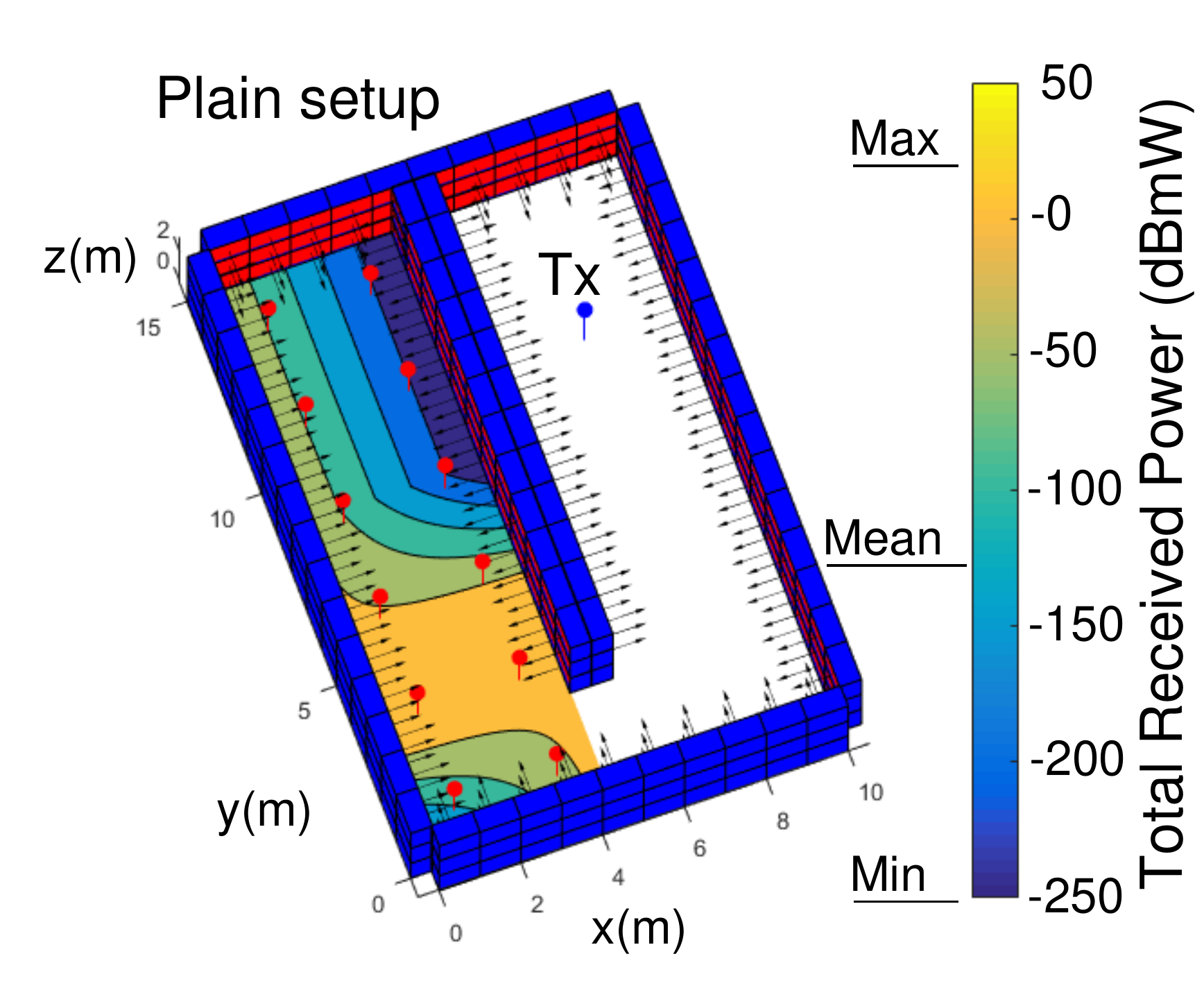}\includegraphics[viewport=60bp 200bp 590bp 650bp,clip,width=0.52\textwidth]{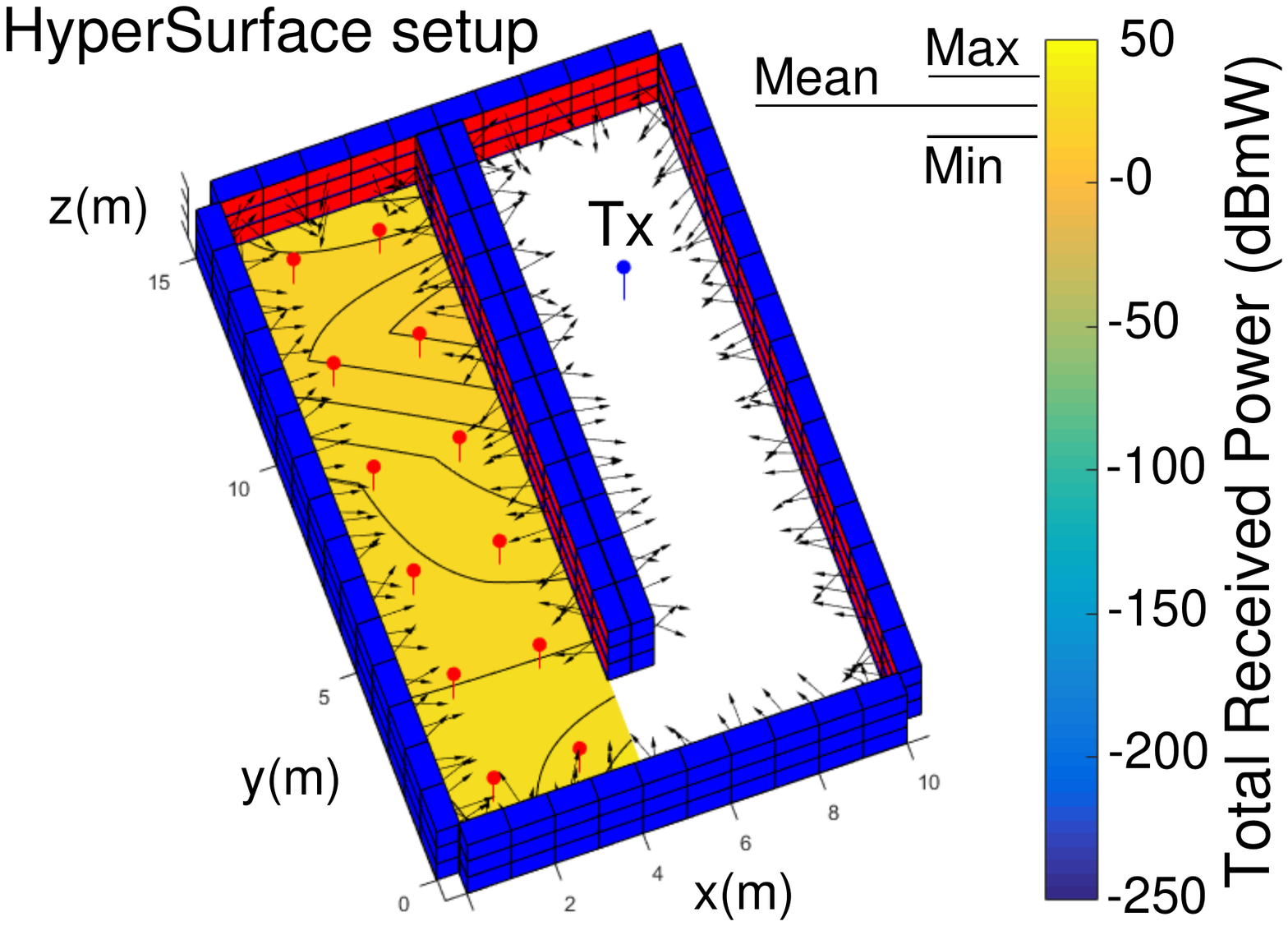}
\par\end{centering}
\caption{\label{fig:maxminP60}Wireless environment optimization case study
(A) for $60\,GHz$ and comparison to the plain case (non-HyperSurface).
The objective is to maximize the minimum total received power over
the NLOS area receivers (red dots). }
\end{figure*}
Existing ray-tracing engines employ common laws of optics to simulate
the propagation of waves. As such, current ray-tracers do not readily
allow for custom wave steering functions. (Absorbing functions, on
the other hand, are readily supported). Thus, to implement steering
functions we work as follows. First, the following observation is
made:
\begin{rem}
Assume a tile and a set of a required wave DoA and a reflection direction
upon it, not abiding by Snell's law. There exists a rotation of the
tile in 3D space that makes the wave DoA and reflection direction
comply with Snell's law.
\end{rem}
Based on this Remark, the custom steering functions are implemented
by tuning the tile's spatial derivative as follows. Since a tile is
a flat and square surface in a 3D space, its spatial derivative is
normally an arrow perpendicular to the tile surface. In order to allow
for custom EM wave steering within the ray-tracing engine, we allow
for virtually rotating the spatial derivative (but not the tile itself)
by proper azimuth and elevation angles. The modified spatial derivative
is then used in all ray-tracing calculations.

The external service is considered to know the tile specifications,
i.e., the tile configuration that corresponds to each virtual angle
combination. The service has obtained the direction of the impinging
wave at each tile via the distributed sensing elements. Subsequently,
it deploys the corresponding STEER or ABSORB commands at each tile,
by applying the corresponding tile configuration.

An EM Tx is placed at position $\left\{ 7,\,12,\,2\right\} \,m$ (with
respect to the origin placed on the floor level, at the upper-left
corner of Fig.~\ref{fig:WirCommExampleNegAngle}). It is equipped
with a half-dipole antenna and transmits at a carrier frequency of
$60\,GHz$ or $2.4\,GHz$ (two studies) and $25\,MHz$ bandwidth.
The transmission power is set to $100\,dBmW$, a high number chosen
to ensure that no propagation paths are disregarded by the ray-tracer
due to its internal, minimum-allowed path loss threshold. The NLOS
area is defined as $x\in\left[0,\,4\right]\,m$, $y\in\left[0,\,15\right]\,m$
and a constant height of $z=1.5\,m$. Within the NLOS area, a set
of $12$ receivers\textendash with antennas identical to the transmitter\textendash are
placed at a regular $2\times6$ uniform grid deployment, with $2.5\,m$
spacing. The receiver grid is centered in the NLOS area. Intermediate
signal reception values, used only for illustration purposes in the
ensuing Figures, are produced by means of interpolation.

\begin{figure}[t]
\begin{centering}
\includegraphics[viewport=40bp 170bp 1000bp 650bp,clip,width=1\columnwidth]{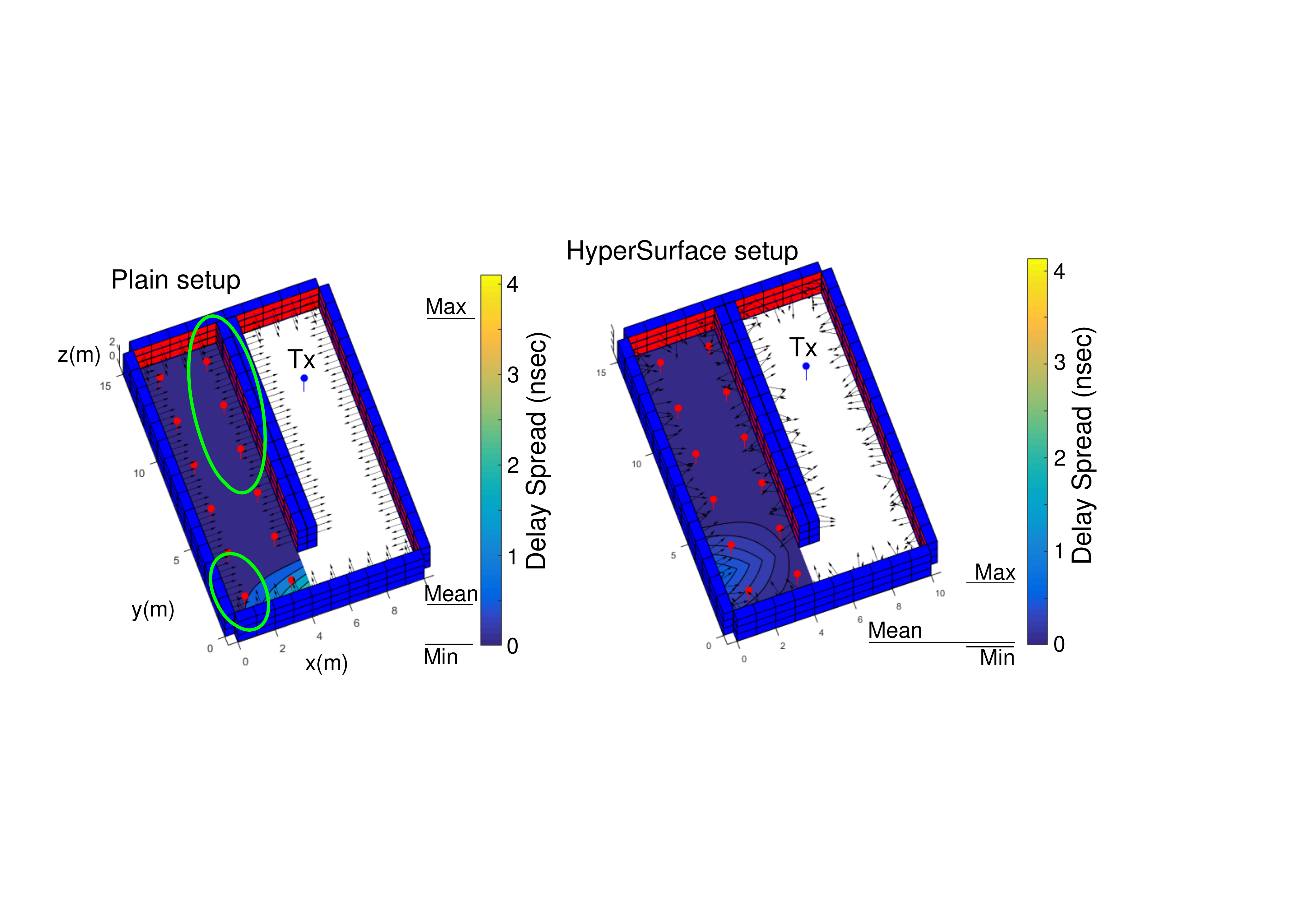}
\par\end{centering}
\caption{\label{fig:minmaxDS60}Wireless environment optimization case study
(B) for $60\,GHz$. The objective is to minimize the maximum delay
spread over the NLOS area, while ensuring a minimum of $1\,dBmW$
total received power per receiver. The circled parts of the plain
setup correspond to disconnected areas. (cf. Fig.~\ref{fig:maxminP60}-left).}
\end{figure}

The evaluation scenario considers two case studies, corresponding
to the path loss and multi-path fading mitigation objectives. In each
case, the state of each of the $222$ tiles is treated as an input
variable of an appropriate objective function which must be optimized.
Given the vastness and discontinuity of the solution space (i.e.,~$222^{26}$
possible tile configurations, positioned at different walls) and the
discrete nature of the input variables, a Genetic Algorithm (GA) is
chosen as the optimization heuristic~\citep{haupt2007genetic}, using
the MATLAB Optimization Toolbox implementation~\citep{MatlabGA}.
GAs are heuristics that are inspired by evolutionary biology principles.
They treat the variables of an optimization problem as \emph{genomes}
which compete with each other in terms of best fitness to an optimization
objective. Good solutions are combined iteratively by exchanging \emph{genes},
i.e., variable sub-parts, producing new generations of solutions.
In the problem at hand, a genome represents a complete tile configuration,
i.e, an array containing the state of the $222$ tiles. A gene represents
the state of each tile, i.e., the specific array elements. Two optimization
cases are studied, denoted as (A) and (B), both for $60\,GHz$ and
$2.4\,GHz$. These are defined as follows:
\begin{itemize}
\item \textbf{\uline{Case study (A)}}. This case expresses the path loss
mitigation goal, and is defined as the following optimization objective:
\emph{Define the optimal tile configurations that maximize the minimum
received power over the 12 receivers in the NLOS area}.
\item \textbf{\uline{Case study (B)}}. The case expresses the multi-path
fading mitigation goal and is defined as the following optimization
objective: \emph{Define the optimal tile configurations that minimize
the maximum delay spread over the 12 receivers in the NLOS area, with
the constraint of ensuring a minimum total received power (custom
threshold}).
\end{itemize}
For Case (B), the thresholds are set to \emph{$1\,dBmW$} for $60\,GHz$,
and \emph{$30\,dBmW$} for $2.4\,GHz$, based on the floor-plan dimensions
and the path loss levels discussed in Section~\ref{sec:App}.
\begin{table}
\centering{}\caption{\textsc{\label{tab:T60}Comparison of total received power (case A)
and power delay profile (case B) with and without HyperSurface (HSF)
Tiles at $60\,GHz$.}}
\begin{tabular}{|c|c|c|c|c|}
\cline{2-5}
\multicolumn{1}{c|}{} & \multicolumn{2}{c|}{Case A ($dBmW$)} & \multicolumn{2}{c|}{Case B ($nsec$)}\tabularnewline
\cline{2-5}
\multicolumn{1}{c|}{} & HSF setup  & Plain setup  & HSF setup  & Plain setup\tabularnewline
\hline
Max  & \textbf{$34.98$}  & $22.63$  & $0.69$  & $3.6$\tabularnewline
\hline
Mean  & $25.38$  & $-75$  & $0.0068$  & $0.48$\tabularnewline
\hline
Min  & $16.13$  & $-250$  & $0.0045$  & $0.007$\tabularnewline
\hline
\end{tabular}
\end{table}
The results for the $60\,GHz$ case are shown in Fig.~\ref{fig:maxminP60},
\ref{fig:minmaxDS60} and are summarized in Table.~\ref{tab:T60}.
Figure~\ref{fig:maxminP60} presents case (A) for the plain (left)
and HyperSurface-enabled (right) environments. In the plain setup,
the tile spatial derivatives (black arrows) are naturally perpendicular
to the tile surfaces. The average received power over the $12$ NLOS
area receivers is $-75\,dBmW$, while the minimum power is $-250\,dBmW$,
which is the lowest level allowed by the ray-tracing engine. Thus,
the bottom-left and the three top-right receivers of the NLOS area
are essentially disconnected in the plain setup. The maximum total
received power is $22.63\,dBmW$.

The right inset of Fig.~\ref{fig:maxminP60} shows the corresponding
results with the HyperSurface functionality enabled. Notably, the
minimum power level over the NLOS area is $16.13\,dB$, which constitutes
a raise by at least $266.13\,dBmW$ with regard to the plain case.
Moreover, the received power becomes essentially uniform over the
NLOS area, ranging between $16.13$ and $34.98\,dBmW$, with an average
of $25.38\,dBmW$. The tile spatial derivatives exhibit a degree of
directivity towards the previously disconnected area parts (e.g.,
cf. left-most wall). Moreover, the top-and bottom tiles across the
height of the walls tend to focus towards the NLOS area height. The
non-uniformity of the derivatives is in accordance with the nature
of the Genetic Algorithm, which is a very exploratory but not gradient-ascending
optimizer~\citep{Luke.2009}. This means that there exists potential
for an even better optimization result near the Genetic Algorithm-derived
solution.
The case (B) results for $60\,GHz$ are shown in Fig.~\ref{fig:minmaxDS60}.
The objective is to minimize the maximum delay spread over the $12$
NLOS receivers, under the constraint for at least $1\,dBmW$ total
received power per receiver. For the plain setup, shown in the left
inset, we note a maximum delay spread of approximately $3.6\,nsec$.
The $1\,dBmW$ minimum power constraint is of course not satisfied,
as previously shown in Fig.~\ref{fig:maxminP60}-left. The circled
areas correspond to the under-powered/disconnected NLOS area parts.
The minimum and average delay spread over the \emph{connected} areas
only are $7\,psec$ and $0.48\,nsec$ respectively. The HyperSurface-enabled
setup (right inset), achieves $5.21$ times lower maximum delay spread
($0.69\,nsec$) than the plain setup, a minimum of $4.5\,psec$ delay
spread ($1.5$ times lower), and an average of $6.8\,psec$ ($70$
times lower). This significant performance improvement is accompanied
by considerable total power levels, in the range of $\left[7.07,\,16.93\right]\,dBmW$
(average:~$10.64\,dBmW$), fulfilling the optimization constraint
of $1\,dBmW$.
\begin{figure*}[t]
\begin{centering}
\includegraphics[viewport=0bp 130bp 2400bp 720bp,clip,width=1\textwidth]{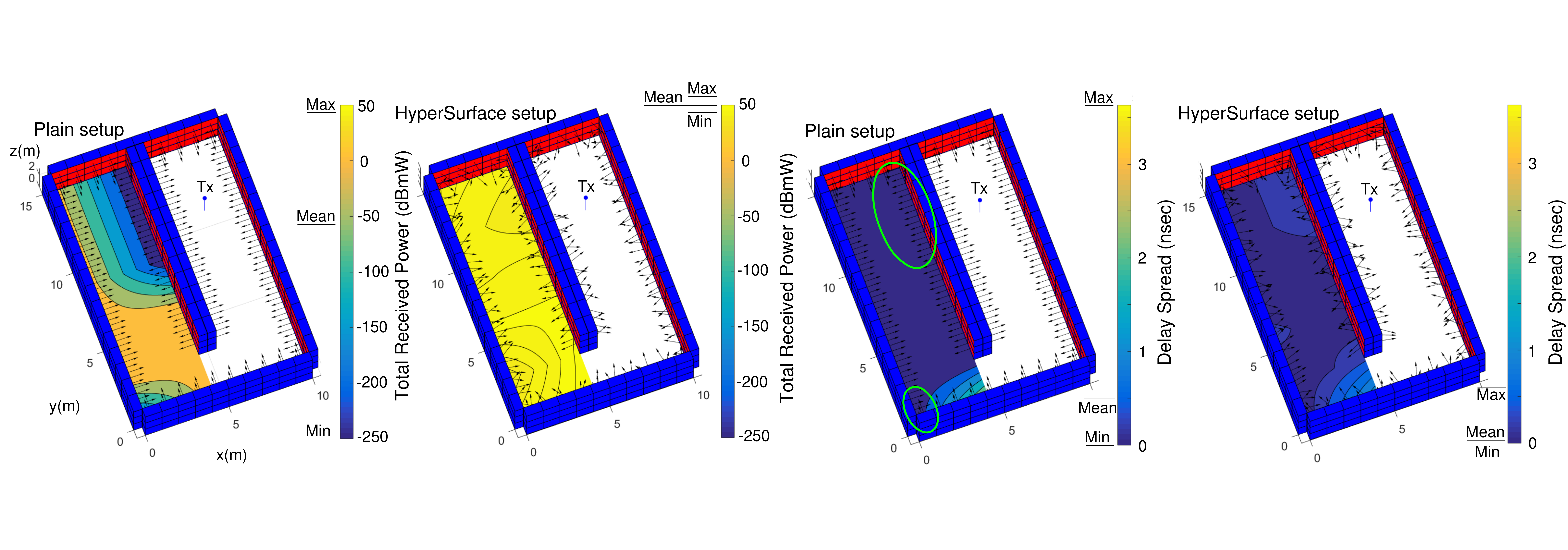}
\par\end{centering}
\caption{\label{fig:all2p4}Wireless environment optimization case studies
(A: left-two insets) and (B: right-two insets) for the $2.4\,GHz$
case. }
\end{figure*}
\begin{table}
\caption{\textsc{\label{tab:T2p4}Comparison of total received power (case
A) and power delay profile (case B) with and without HyperSurface
(HSF) Tiles at $2.4\,GHz$.}}
\centering{}%
\begin{tabular}{|c|c|c|c|c|}
\cline{2-5}
\multicolumn{1}{c|}{} & \multicolumn{2}{c|}{Case A ($dBmW$)} & \multicolumn{2}{c|}{Case B ($nsec$)}\tabularnewline
\cline{2-5}
\multicolumn{1}{c|}{} & HSF setup  & Plain setup  & HSF setup  & Plain setup\tabularnewline
\hline
Max  & \textbf{$59.81$}  & $47$  & $0.68$  & $3.65$\tabularnewline
\hline
Mean  & $51.37$  & $-58$  & $0.067$  & $0.47$\tabularnewline
\hline
Min  & $45.13$  & $-250$  & $0.0029$  & $0.0014$\tabularnewline
\hline
\end{tabular}
\end{table}
The results for the $2.4\,GHz$ case are similar to the $60\,GHz$
in terms of improvement, and are collectively given in Fig.~\ref{fig:all2p4}
and Table~\ref{tab:T2p4}. The objective in the two leftmost panels
is to maximize the minimum total received power over the $12$ receivers
in the NLOS area. The plain setup achieves $-250$, $-58$ and $47\,dBmW$
minimum, average and maximum total received power, respectively. The
HyperSurface setup yields considerably improved results, with $45.13$,
$51.37$ and $59.81\,dBmW$ minimum, average and maximum total received
power, respectively. Thus, there is a gain of $295.13\,dBmW$ in minimum
received power.

The delay spread improvement is also significant, as shown in the
two rightmost panels. The plain setup yields $1.4\,psec$, $0.47\,nsec$
and $3.65\,nsec$ minimum, average and maximum delay spread values,
with $4$ disconnected receivers (circled parts, cf. first inset of
Fig.~\ref{fig:all2p4}). The corresponding HyperSurface-enabled setup
achieves $2.9\,psec$, $67\,psec$ and $0.68\,nsec$ min/average/max
respectively. Moreover, it ensures a minimum total received power
of $34.12\,dBmW$, successfully meeting the $30\,dBmW$ optimization
constraint.

\subsection{Security objectives}

\begin{figure}[t]
\begin{centering}
\includegraphics[width=0.8\columnwidth]{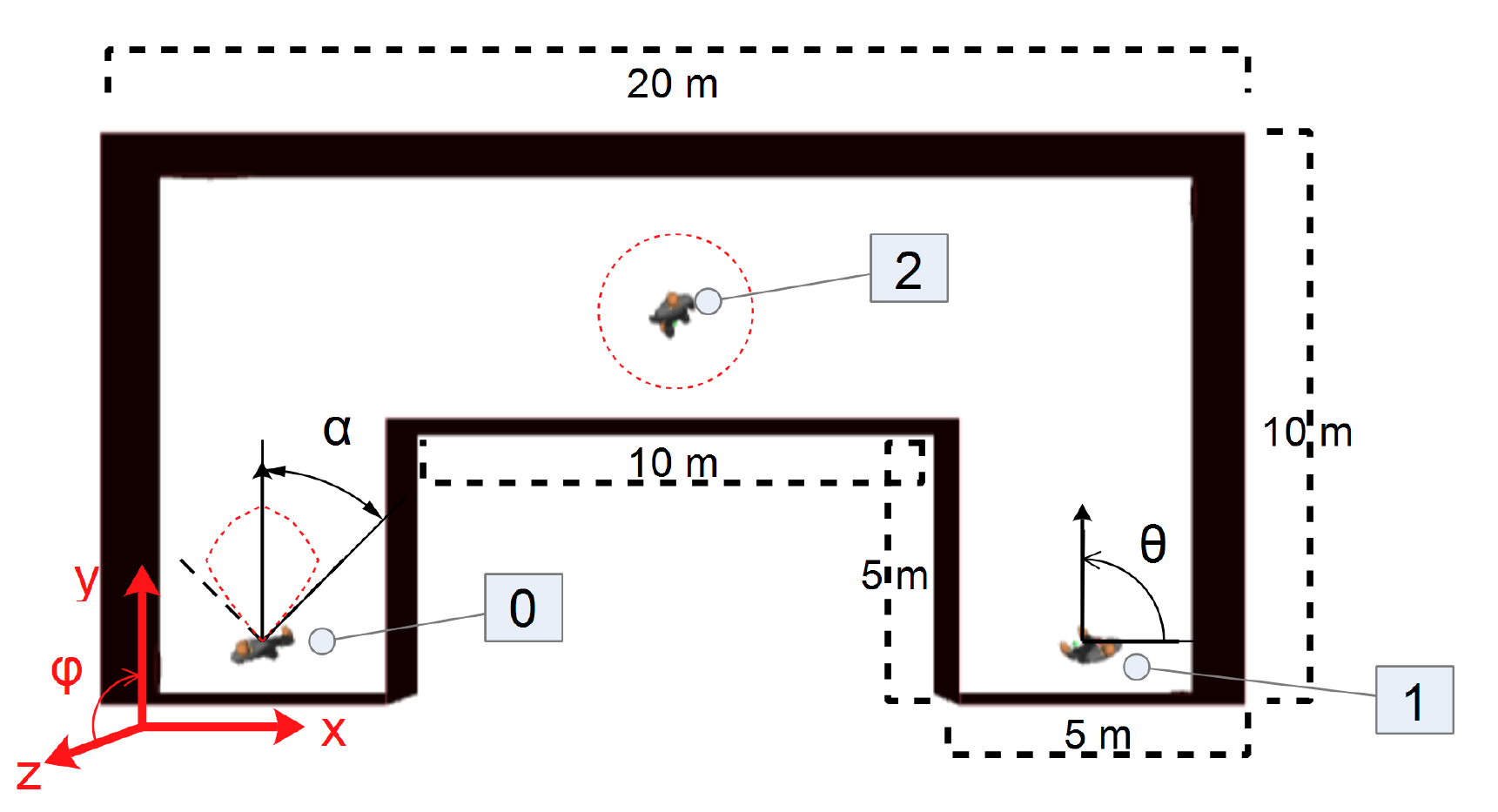}
\par\end{centering}
\caption{\label{fig:SECSETUP}Setup for the eavesdropping mitigation evaluation
scenario. User $0$ seeks to send data to user $1$, and user $2$
acts as the eavesdropper.}
\end{figure}

\begin{figure}[t]
\begin{centering}
\includegraphics[viewport=0bp 0bp 274bp 254bp,width=0.48\columnwidth]{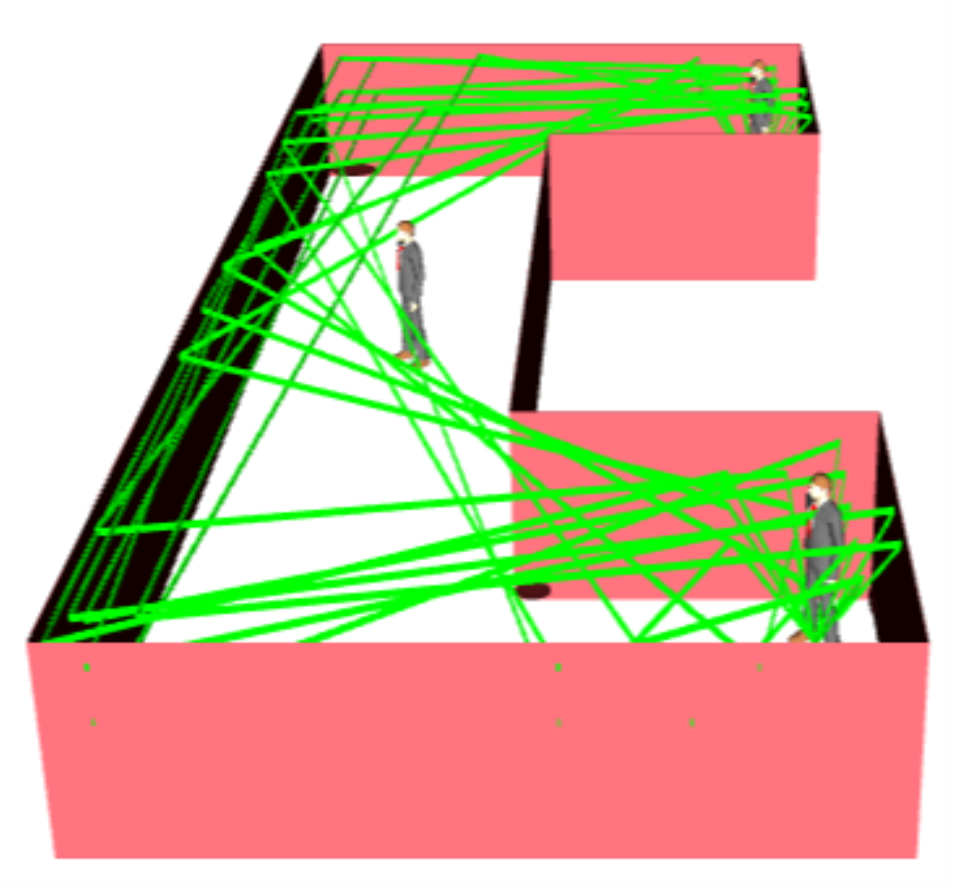}
\includegraphics[viewport=0bp 0bp 409bp 210bp,width=0.48\columnwidth]{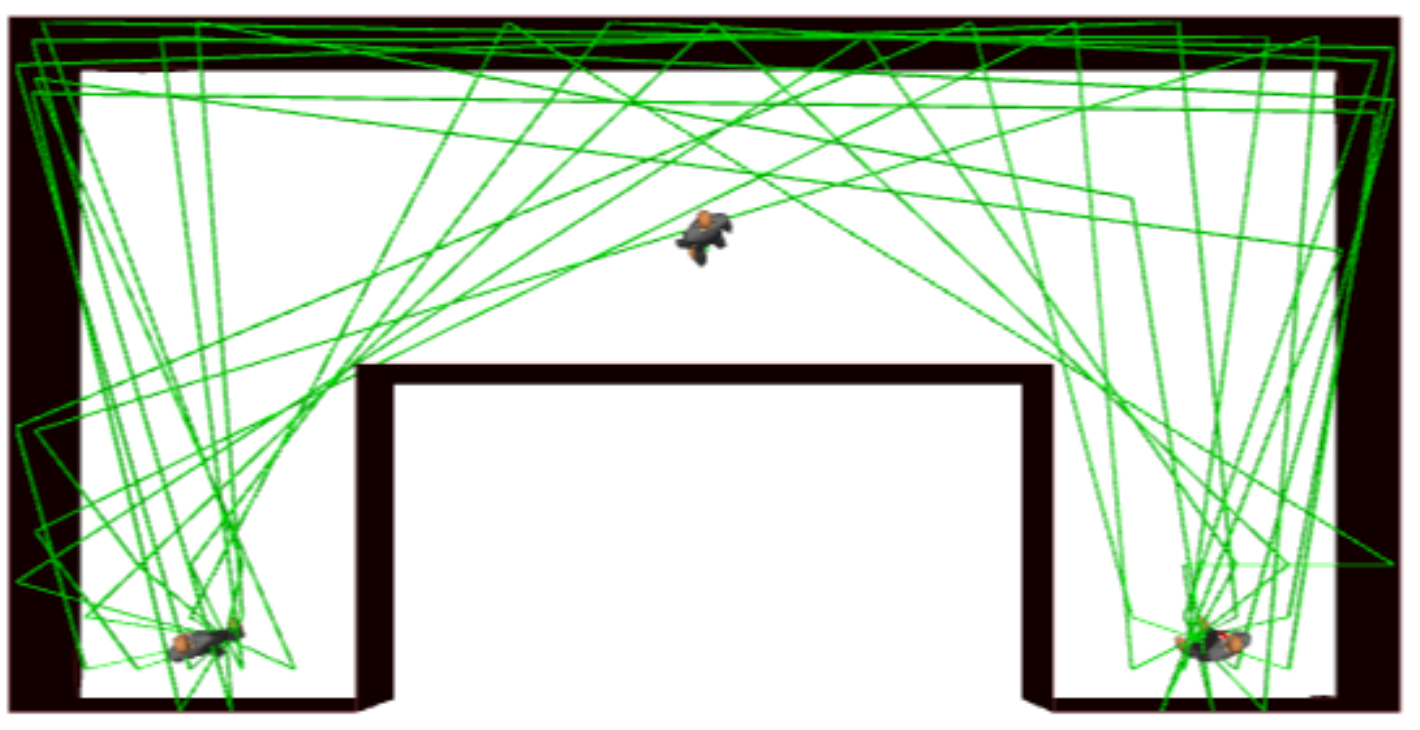}
\par\end{centering}
\caption{\label{fig:HSFPROP}HyperSurface-controlled propagation, side and
top views. Propagated waves avoid the eavesdropper by remaining confined
at the highest parts of the floorplan.}
\end{figure}

We proceed to evaluate the eavesdropping mitigation approaches described
in Section~\ref{subsec:Physical-Layer-Security-Objectiv}, i.e.,
avoid an eavesdropper by routing rays away from him, or by tuning
the phase of rays in order to cancel each other out near the eavesdropper.
To this end, we consider the setup of Fig.~\ref{fig:SECSETUP}. Two
users, a transmitter (user 0) and a receiver (user 1) are placed in
a NLOS setting, with an eavesdropper (user 2) located in-between.
Table~\ref{tab:TSECParams} describes the parameters of the setup.
The following notes are made:
\begin{itemize}
\item The assumed programmable wireless environment deployment is partial.
Only the ceiling and the highest part of the walls are covered with
tiles. This approach provides cost and deployment advantages. First,
fewer tiles naturally translate to lower cost and overall complexity.
Furthermore, ceilings and upper parts of walls are commonly vacant
of other use, while offering easy access to power supply (e.g., via
the lights power lines).
\item The tiles are considered to attenuate impinging waves by a constant
factor of 1~\% of the carried power, which constitutes a typical
efficiency index for state-of-the-art metasurfaces~\citep{Banerjee.2011}.
\item The considered tiles functionalities include collimation and carrier
phase control~\citep{Chen.2016}. Collimation is the effect of aligning
EM waves to propagate over a flat front, rather than to dissipate
over an ever-growing sphere. Thus, the path loss between two tiles
is not subject to the $\propto\nicefrac{1}{d^{2}}$ rule, $d$ being
their distance. This rule is only valid for the first impact, i.e.,
from the transmitter to its LOS tiles. The antenna aperture and gains
are taken into account as usual. The phase control is required only
for the corresponding physical later security approach, described
in Section~\ref{subsec:Physical-Layer-Security-Objectiv}.
\item The antenna patterns of the transmitter and the receiver are simplified
as single-lobe sinusoids, with the characteristics and $\theta,\,\phi$
orientiation shown in Fig.~\ref{fig:SECSETUP} and Table~\ref{tab:TSECParams}.
In one scenario, we assume that the mobile devices have beamforming
capabilites and are able to turn the antenna lobe towards the ceiling,
in conjunction with the mobile devices' gyroscopes.
\item The eavesdropper's antenna is considered to be isotropic. Moreover,
all rays pass through him unobstructed. In contrast, the bodies of
the users are modeled as spheres of radius $0.5$~m, fully blocking
impinging waves.
\begin{table}
\centering{}\caption{\textsc{\label{tab:TSECParams}Simulation parameters / security objectives.}}
\begin{tabular}{|c|c|}
\hline
User 0 position & \textbf{$x:\,2.5,\,y:\,1,\,z:\,1\,m$} \tabularnewline
\hline
User 1 position & \textbf{$x:\,17.5,\,y:\,1,\,z:\,1\,m$} \tabularnewline
\hline
User 2 position & \textbf{$x:\,10,\,y:\,7,\,z:\,1\,m$} \tabularnewline
\hline
Ceiling Height & $3$~$m$\tabularnewline
\hline
Tile Dimensions  & \textbf{$75\times75\,cm$} \tabularnewline
\hline
Tile Placement & Ceiling, Upper part of walls ($>1.5\,\text{m}$)\tabularnewline
\hline
Tile Power loss & $1$~\%~per ray bounce\tabularnewline
\hline
Tile Functions & COLLIMATE, STEER, PHASE\_ALTER\tabularnewline
\hline
Frequency & $2.4\,GHz$ \tabularnewline
\hline
Tx Power (User 0) & $-30\,dBm$\tabularnewline
\hline
\multirow{2}{*}{Antenna types} & \multirow{1}{*}{Users 0, 1~Single lobe sinusoid ($\alpha=30^{o}$)}\tabularnewline
\cline{2-2}
 & User 2 (Eavesdropper)~$\to$isotropic\tabularnewline
\hline
\multirow{2}{*}{Antenna orientation} & Fig~\ref{fig:FREEPROP}, Users 0, 1~$:\left(\theta=90^{o},\,\phi=90^{o}\right)$\tabularnewline
\cline{2-2}
 & Fig~\ref{fig:HSFPROP}, Users 0, 1~$:\left(\theta=90^{o},\,\phi=0^{o}\right)$\tabularnewline
\hline
Max ray bounces & $50$\tabularnewline
\hline
Min considered ray power & -$250$~dBm\tabularnewline
\hline
\end{tabular}
\end{table}
\begin{figure}[t]
\begin{centering}
\includegraphics[viewport=0bp 0bp 293bp 259bp,width=0.48\columnwidth]{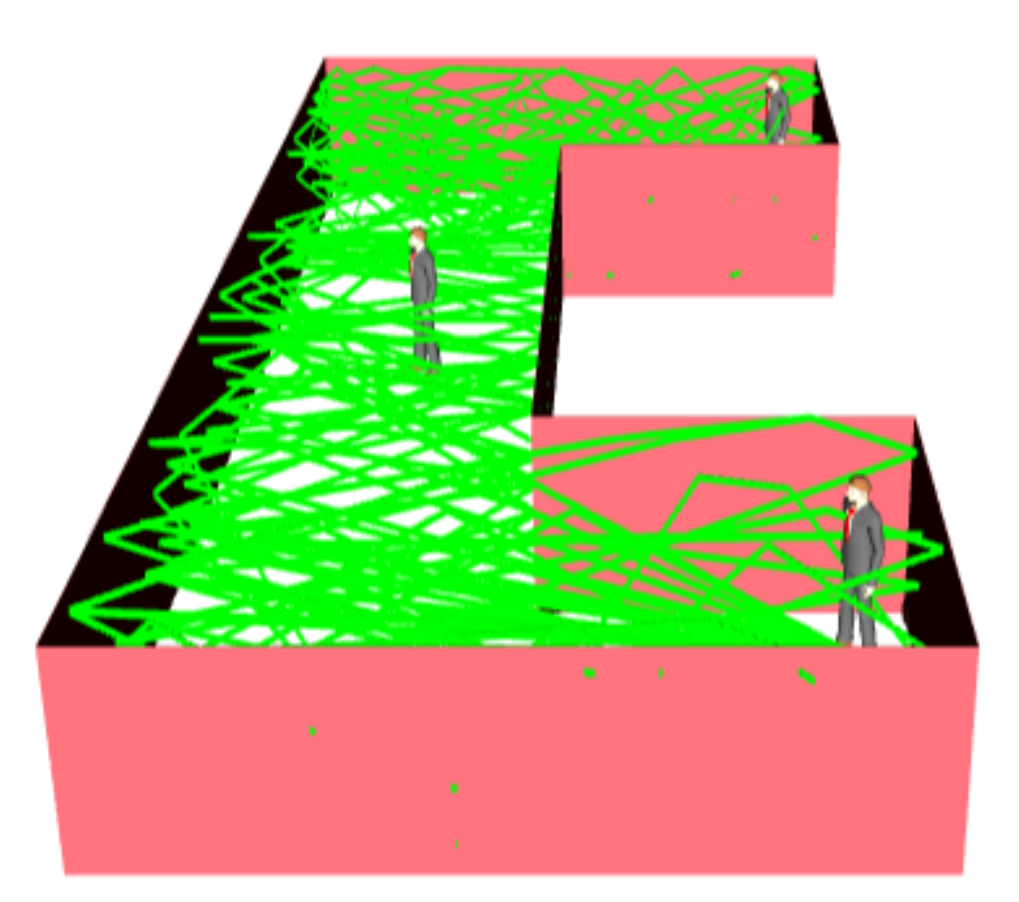}
\includegraphics[viewport=0bp 0bp 406bp 208bp,width=0.48\columnwidth]{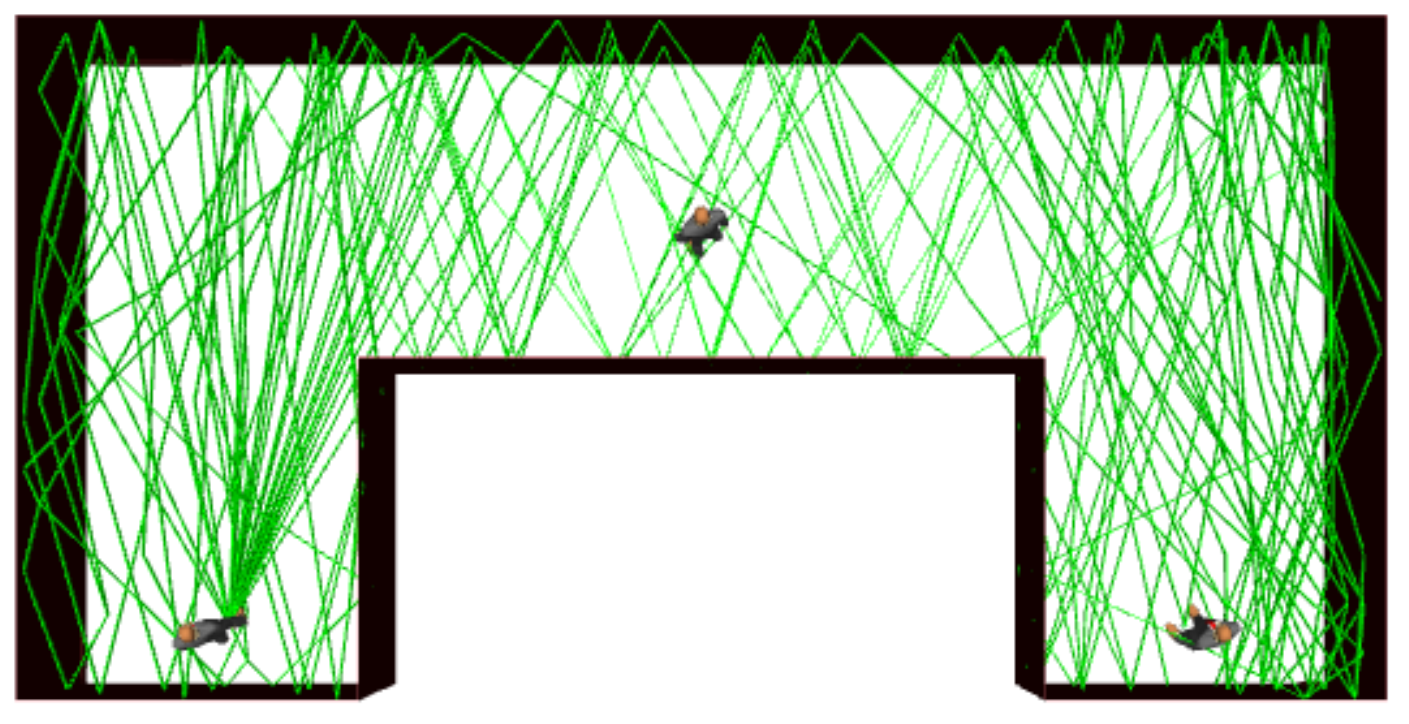}
\par\end{centering}
\caption{\label{fig:FREEPROP}Natural propagation (without HyperSurfaces),
side and top views.}
\end{figure}
\end{itemize}
Figure~\ref{fig:FREEPROP} shows the natural propagation, in an environment
without HyperSurfaces. The propagation is expectedly chaotic, while
several rays are visibly intercepted by the eavesdropper. On the other
hand, the propagation in the case of programmable wireless environments
is more well-defined, as shown in Fig.~\ref{fig:HSFPROP}. In this
case, the user devices have employed beamforming to turn their antenna
lobes towards the ceiling. COLLIMATE and STEER functions are applied
to tiles, routing rays from the transmitter to the receiver. This
is attained by first calculating the K tile-disjoint paths from the
transmitter to the receiver, and then deploying STEER functions accordingly.
The first impact tiles are configured to additionally collimate impinging
waves as described.
The performance and security benefits of the programmable wireless
environment as summarized in Table~\ref{tab:TsecPOWnoPHASE}. In
the plain setup, the intended user receives -$83\,dBm$ total power,
i.e., a total path loss of $53\,dB$. Moreover, the eavesdropper has
better reception quality than the intended receiver, since he is physically
closer to the transmitter. In the programmable environment case, the
eavesdropping is completely mitigated, since the EM propagation remains
confined to the upper part of the floorplan. Moreover, the intended
recipient has a considerably better reception of $-47\,dBm$, i.e.,
a $36\,dB$ improvement over the plain case.

\begin{table}
\centering{}\caption{\textsc{\label{tab:TsecPOWnoPHASE}Comparison of total received/eavesdropped
power, without 'phase\_alter'.}}
\begin{tabular}{|c|c|c|}
\cline{2-3}
\multicolumn{1}{c|}{} & \multicolumn{2}{c|}{Received Power ($dBmW$)}\tabularnewline
\cline{2-3}
\multicolumn{1}{c|}{} & HSF setup (Fig.~\ref{fig:HSFPROP}) & Plain setup (Fig.~\ref{fig:FREEPROP}) \tabularnewline
\hline
User 1 & \textbf{$-47$}  & $-83$ \tabularnewline
\hline
User 2 & \multirow{2}{*}{$-\infty$ } & \multirow{2}{*}{$-76$ }\tabularnewline
(eavesdropper) &  & \tabularnewline
\hline
\end{tabular}
\end{table}
Finally, we proceed to evaluate the phase control-based approach for
eavesdropping mitigation. To this end, we return to the plain propagation
shown in Fig.~\ref{fig:FREEPROP}, i.e., no STEER or COLLIMATE functions
are applied. Instead PHASE\_ALTER functions are exclusively applied,
to achieve the effect described in the context of Fig.~\ref{fig:PhaseControlApproach}.

\begin{figure}
\begin{centering}
\subfloat[\label{fig:Without-'PHASE_ALTER',-received}Without 'PHASE\_ALTER',
received power $-76\,dBm$.]{\begin{centering}
\includegraphics[viewport=100bp 270bp 500bp 520bp,clip,width=0.47\columnwidth]{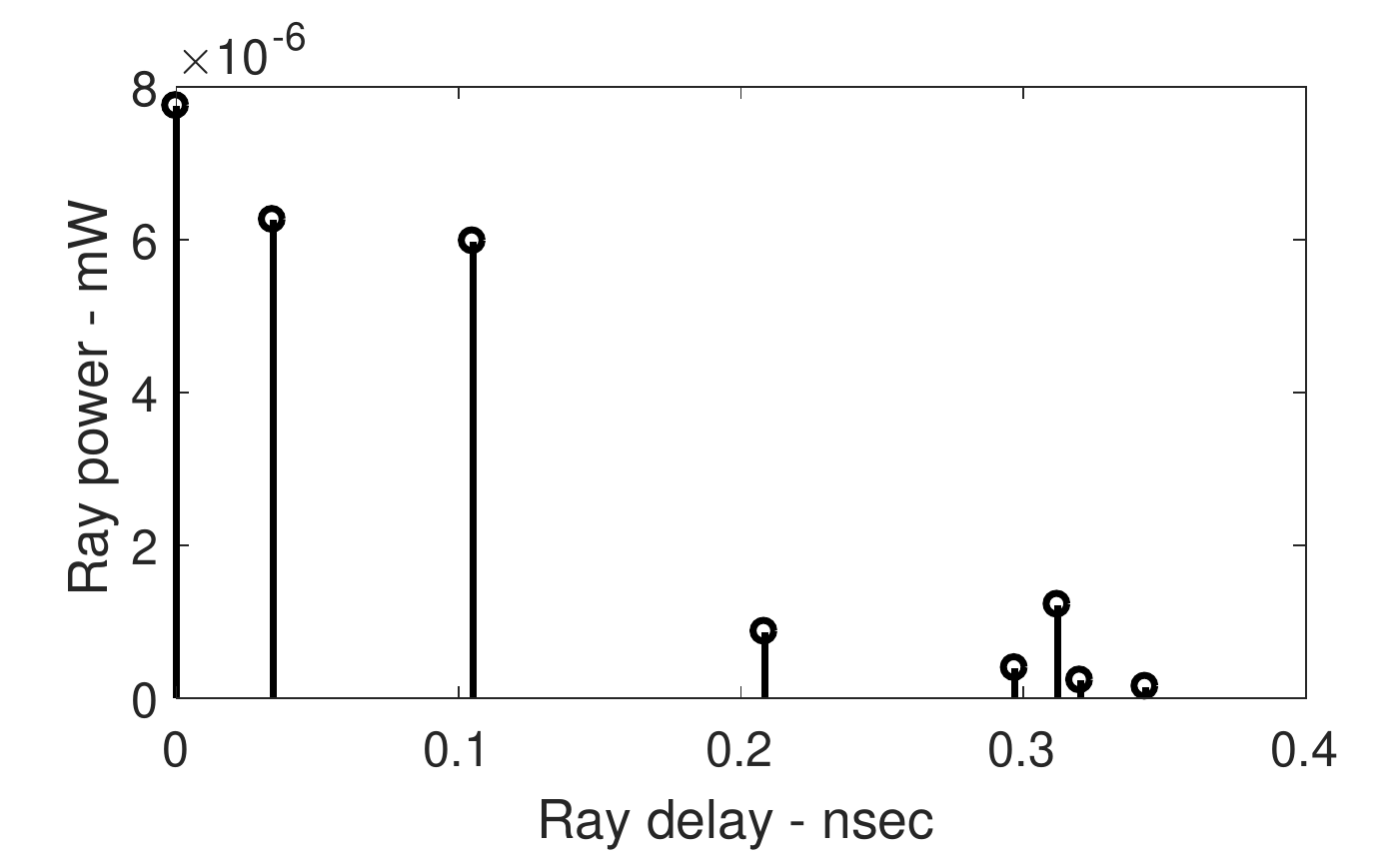}
\par\end{centering}
}~~\subfloat[\label{fig:With-'PHASE_ALTER',-received}With 'PHASE\_ALTER', received
power $-82.7\,dBm$.]{\begin{centering}
\includegraphics[viewport=100bp 270bp 500bp 520bp,clip,width=0.47\columnwidth]{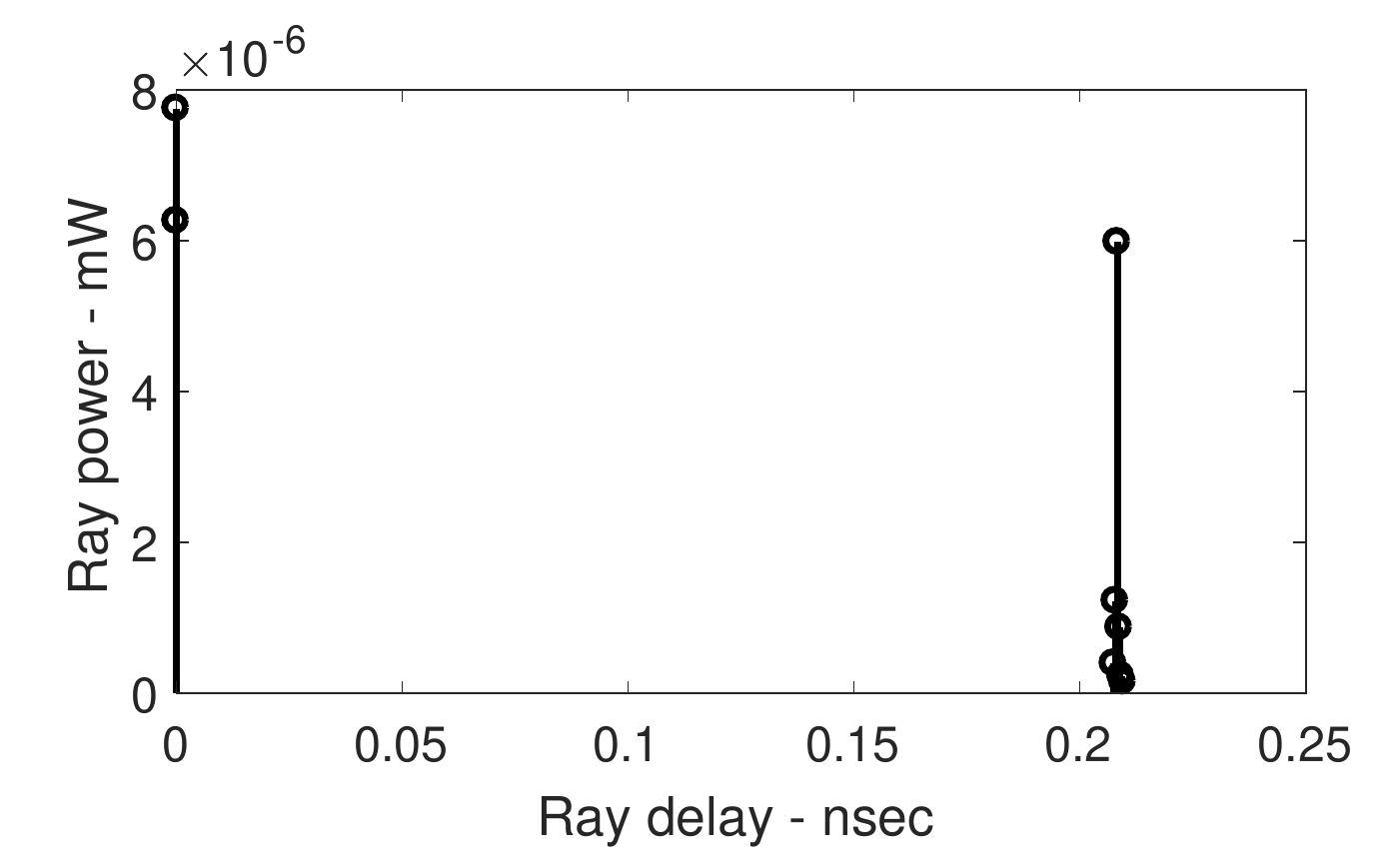}
\par\end{centering}
}
\par\end{centering}
\caption{Power Delay Profile for user 2 under natural propagation (Fig.~\ref{fig:FREEPROP}),
without and with 'PHASE\_ALTER'.}
\end{figure}
Figure~\ref{fig:Without-'PHASE_ALTER',-received} shows the power-delay
profile of the eavesdropper in the plain case. The x-axis is the timing
of each received ray relative to the earlier ray, modulo the carrier
period (i.e., multiplication of the x-axis values by $2\pi f$ return
the relative phase in radians). The phase control-based mitigation
pertains to altering the phase of each ray in a manner that cancels
them out. In the case of Fig.~\ref{fig:Without-'PHASE_ALTER',-received},
the second and third rays are carry almost the same power and, therefore,
they can be moved to opposite phases to cancel each other out. Rays
four to eight can be synchronized and be opposed to the first and
strongest ray. The described effects are shown in Fig.~\ref{fig:With-'PHASE_ALTER',-received}.

The described phase control offers almost $6\,dB$ of extra signal
attenuation at the eavesdropper. The maximal gains of this approach
is strongly depended on the original power delay profile. The effects
are expected to be optimal when there exist group rays that carry
approximately the same power. Such groups can be subsequently phase-controlled
to cancel each other out. Finally, it is noted that phase control
is naturally expected to be sensitive to errors, such as the perceived
location of users. Adaptive control loops are expected to be required
in order to achieve the eavesdropping mitigation gains. Moreover,
the phase control approach may only be used as a last resort, i.e.,
when the ray-routing approach cannot offer routes that avoid the potential
eavesdropper.

\subsection{Discussion and Research Directions\label{sec:future}}

The results of Section~\ref{sec:Evaluation} demonstrated the performance
and security potential of the proposed softwarization of wireless
indoor environments.

In terms of performance objectives, the evaluation showcased path
loss and multi-path fading mitigation. Even at the highly-challenging
$60\,GHz$ communications, a HyperSurface tile-coated indoor setup
exhibited significant improvements in received power levels and delay
spread. Such traits can benefit the communication distance of devices
and their energy consumption, without dissipating more energy in the\textendash already
EM-strained\textendash environments via retransmitters. In terms of
security, HyperSurfaces are shown to be promising enforcers of Physical
Layer Security, due to their ability to micro-manage EM waves. Strong
protection against eavesdropping can be attained by routing waves
via improbable paths, avoiding potential eavesdropper altogether,
or by ensuring coherent reception only near the receiver.

These promising traits can encourage further exploration of the HyperSurface
concept in additional usage domains. Multiple applications can be
studied in both indoor and outdoor environments, and in the context
of multiple systems, such as 5G, IoT and D2D, where security, ultra-low
latency, high bandwidth, and support for massive numbers of devices
are important~\citep{Aijaz.2017}. Moreover, HyperSurfaces may act
as an enabler for upcoming $THz$ communications. Operation in this
band promises exceptional data rates and hardware size minimization
at the nano-level, which can enable a wide range of groundbreaking
applications~\citep{akyildiz2015internet}. Nonetheless, the $THz$
band is susceptible to acute signal attenuation owed to molecular
absorption. HyperSurfaces with graphene-based meta-atom designs could
act as a smart environment for up to $1.8\,THz$ communications~\citep{Fan.2013},
mitigating the attenuation effects and extending the communication
range.

Further research directions can also study the placement and topology
of tiles within an environment, as well as their networking aspects.
Partial, optimized coverage of an environment may provide sufficient
performance gains over a full deployment. Coverage versus gains studies
can quantify this aspect. Moreover, the networking topology, e.g.,
comprising a hierarchy of environment controllers and their corresponding
sets of tiles is subject to optimization, in order to achieve a timely
EM wave sensing and environment re-configuration loop. Communication
protocols between tiles and controllers is similarly subject to optimization.

Finally, wireless power transfer is another promising application,
apart from wireless performance and security. The ability of HyperSurfaces
to collimate, steer and focus EM waves can be employed to transfer
and charge devices wirelessly over long distances.

\section{Conclusion\label{sec:Conclusion}}

In this paper we proposed an indoor wireless communication paradigm
where the electromagnetic propagation environment becomes aware of
the ongoing communications within it.The key idea is to coat objects
such as walls, doors and furniture with HyperSurface tiles, a forthcoming
type of material with programmable electromagnetic behavior. HyperSurfaces
can exert fine-grained control over impinging electromagnetic waves,
steering them toward completely custom directions, polarizing them
or fully absorbing them. HyperSurfaces have inter-networking capabilities,
allowing for the first time the participation of electromagnetic properties
of materials into control loops. A central server maintains a view
of the communicating devices within an indoor space, and subsequently
sets the tile electromagnetic configuration in accordance with any
optimization objective. The HyperSurface tile concept has been evaluated
in $2.4\,\text{and}\,60\,GHz$ setups, which demonstrated its high
potential for path loss and multi-path fading mitigation, from microwave
to mm-wave setups. Moreover, HyperSurfaces were shown to be efficient
enforcers of physical layer security, micromanaging the propagation
of electromagnetic waves in novel, eavesdropping-blocking ways.

\section*{Acknowledgment}

This work was partially funded by the European Union via the Horizon
2020: Future Emerging Topics call (FETOPEN), grant EU736876, project
VISORSURF (http://www.visorsurf.eu).

\bibliographystyle{elsarticle-num}

\begin{thebibliography}{10}

\bibitem{pi2016millimeter}
Z.~Pi, J.~Choi, and R.~Heath, ``{Millimeter-wave gigabit broadband evolution
  toward 5G: fixed access and backhaul},'' {\em {IEEE Communications
  Magazine}}, vol.~54, no.~4, pp.~138--144, 2016.

\bibitem{yilmaz2016millimetre}
T.~Yilmaz and O.~B. Akan, ``{Millimetre-Wave Communications for 5G Wireless
  Networks},'' {\em {Opportunities in 5G Networks: A Research and Development
  Perspective}}, pp.~425--440, 2016.

\bibitem{Aijaz.2017}
A.~Aijaz, M.~Simsek, M.~Dohler, and G.~Fettweis, ``{Shaping 5G for the Tactile
  Internet},'' in {\em {5G Mobile Comm.}}, pp.~677--691, Springer, 2017.

\bibitem{kelif20163d}
J.-M. Kelif {\em et~al.}, ``{A 3D beamforming analytical model for 5G wireless
  networks},'' in {\em {14th WiOpt}}, pp.~1--8, 2016.

\bibitem{REFLECTARRAYS}
S.~V. Hum, M.~Okoniewski, and R.~J. Davies, ``Modeling and design of
  electronically tunable reflectarrays,'' {\em IEEE transactions on Antennas
  and Propagation}, vol.~55, no.~8, pp.~2200--2210, 2007.

\bibitem{huang2017multi}
J.~Huang, L.~I. Wenxiang, Y.~Su, and F.~Wang, ``{Multi-rate combination of
  partial information based routing and adaptive modulation and coding for
  space deterministic delay/disruption tolerant networks},'' {\em {IET
  Communications}}, 2017.

\bibitem{reflectInfocom.2017}
S.~Han and K.~G. Shin, ``{Enhancing Wireless Performance Using Reflectors},''
  in {\em {INFOCOM 2017}}, pp.~1--10.

\bibitem{chen2017promoting}
Y.~Chen, S.~He, F.~Hou, Z.~Shi, and J.~Chen, ``{Promoting device-to-device
  communication in cellular networks by contract-based incentive mechanisms},''
  {\em {IEEE Network}}, vol.~31, no.~3, pp.~14--20, 2017.

\bibitem{Zhu.2017}
A.~Y. Zhu {\em et~al.}, ``{Traditional and emerging materials for optical
  metasurfaces},'' {\em {Nanophotonics}}, vol.~6, no.~2, 2017.

\bibitem{Minovich.2015}
A.~E. Minovich, A.~E. Miroshnichenko, A.~Y. Bykov, T.~V. Murzina, D.~N. Neshev,
  and Y.~S. Kivshar, ``{Functional and nonlinear optical metasurfaces: Optical
  metasurfaces},'' {\em {Laser {\&} Photonics Reviews}}, vol.~9, no.~2,
  pp.~195--213, 2015.

\bibitem{Lucyszyn.2010}
S.~Lucyszyn, {\em {Advanced RF MEMS}}.
\newblock {The Cambridge RF and microwave engineering series}, NY: Cambridge
  Univ. Press, 2010.

\bibitem{Chen.2016}
H.-T. Chen, A.~J. Taylor, and N.~Yu, ``{A review of metasurfaces: physics and
  applications},'' {\em {Reports on progress in physics. Physical Society
  (Great Britain)}}, vol.~79, no.~7, p.~076401, 2016.

\bibitem{Lee.2012}
S.~H. Lee {\em et~al.}, ``{Switching terahertz waves with gate-controlled
  active graphene metamaterials},'' {\em {Nature Materials}}, vol.~11, no.~11,
  pp.~936--941, 2012.

\bibitem{wifihome}
``Wi-fi home design.''

\bibitem{yu2011light}
N.~Yu, P.~Genevet, M.~A. Kats, F.~Aieta, J.-P. Tetienne, F.~Capasso, and
  Z.~Gaburro, ``Light propagation with phase discontinuities: generalized laws
  of reflection and refraction,'' {\em science}, p.~1210713, 2011.

\bibitem{hotnetspapper}
A.~Welkie, L.~Shangguan, J.~Gummeson, W.~Hu, and K.~Jamieson, ``Programmable
  radio environments for smart spaces,'' in {\em Proceedings of the 16th ACM
  Workshop on Hot Topics in Networks}, HotNets-XVI, (New York, NY, USA),
  pp.~36--42, ACM, 2017.

\bibitem{Liaskos.2015b}
C.~Liaskos, A.~Tsioliaridou, A.~Pitsillides, I.~F. Akyildiz, N.~Kantartzis,
  A.~Lalas, X.~Dimitropoulos, S.~Ioannidis, M.~Kafesaki, and C.~Soukoulis,
  ``{Design and Development of Software Defined Metamaterials for
  Nanonetworks},'' {\em {IEEE Circuits and Systems Magazine}}, vol.~15, no.~4,
  pp.~12--25, 2015.

\bibitem{Banerjee.2011}
B.~Banerjee, {\em {An introduction to metamaterials and waves in composites}}.
\newblock Boca Raton, FL: CRC Press/Taylor {\&} Francis Group, 2011.

\bibitem{wyner1975wire}
A.~D. Wyner, ``The wire-tap channel,'' {\em Bell system technical journal},
  vol.~54, no.~8, pp.~1355--1387, 1975.

\bibitem{bloch2008wireless}
M.~Bloch, J.~Barros, M.~R. Rodrigues, and S.~W. McLaughlin, ``Wireless
  information-theoretic security,'' {\em IEEE Transactions on Information
  Theory}, vol.~54, no.~6, pp.~2515--2534, 2008.

\bibitem{akyildiz20165g}
I.~F. Akyildiz, S.~Nie, S.-C. Lin, and M.~Chandrasekaran, ``{5G roadmap: 10 key
  enabling technologies},'' {\em Computer Networks}, vol.~106, pp.~17--48,
  2016.

\bibitem{wang2016physical}
C.~Wang and H.~Wang, ``Physical layer security in millimeter wave cellular
  networks,'' {\em IEEE Transactions on Wireless Communications}, vol.~15,
  pp.~5569--5585, Aug 2016.

\bibitem{zhu2016physical}
Y.~Zhu, L.~Wang, K.~Wong, and R.~W. Heath, ``Physical layer security in
  large-scale millimeter wave ad hoc networks,'' in {\em 2016 IEEE Global
  Communications Conference (GLOBECOM)}, pp.~1--6, Dec 2016.

\bibitem{yang2015Safeguarding}
N.~Yang, L.~Wang, G.~Geraci, M.~Elkashlan, J.~Yuan, and M.~D. Renzo,
  ``Safeguarding 5g wireless communication networks using physical layer
  security,'' {\em IEEE Communications Magazine}, vol.~53, pp.~20--27, April
  2015.

\bibitem{Kapetanovic2015physical}
D.~Kapetanovic, G.~Zheng, and F.~Rusek, ``Physical layer security for massive
  mimo: An overview on passive eavesdropping and active attacks,'' {\em IEEE
  Communications Magazine}, vol.~53, pp.~21--27, June 2015.

\bibitem{garnaev2014incorporating}
A.~Garnaev, M.~Baykal-Gursoy, and H.~V. Poor, ``Incorporating attack-type
  uncertainty into network protection,'' {\em IEEE Transactions on Information
  Forensics and Security}, vol.~9, no.~8, pp.~1278--1287, 2014.

\bibitem{wu2016secure}
Y.~Wu, R.~Schober, D.~W.~K. Ng, C.~Xiao, and G.~Caire, ``Secure massive mimo
  transmission with an active eavesdropper,'' {\em IEEE Transactions on
  Information Theory}, vol.~62, pp.~3880--3900, July 2016.

\bibitem{zhang2018pilot}
X.~Zhang and E.~W. Knightly, ``Pilot distortion attack and zero-startup-cost
  detection in massive mimo network: From analysis to experiments,'' {\em IEEE
  Transactions on Information Forensics and Security}, 2018.

\bibitem{zhu2014secure}
J.~Zhu, R.~Schober, and V.~K. Bhargava, ``Secure transmission in multicell
  massive mimo systems,'' {\em IEEE Transactions on Wireless Communications},
  vol.~13, no.~9, pp.~4766--4781, 2014.

\bibitem{haus2017Security}
M.~Haus, M.~Waqas, A.~Y. Ding, Y.~Li, S.~Tarkoma, and J.~Ott, ``Security and
  privacy in device-to-device (d2d) communication: A review,'' {\em IEEE
  Communications Surveys Tutorials}, vol.~19, pp.~1054--1079, Secondquarter
  2017.

\bibitem{harrison2013coding}
W.~K. Harrison, J.~Almeida, M.~R. Bloch, S.~W. McLaughlin, and J.~Barros,
  ``Coding for secrecy: An overview of error-control coding techniques for
  physical-layer security,'' {\em IEEE Signal Processing Magazine}, vol.~30,
  pp.~41--50, Sept 2013.

\bibitem{poor2017wireless}
H.~V. Poor and R.~F. Schaefer, ``Wireless physical layer security,'' {\em
  Proceedings of the National Academy of Sciences}, vol.~114, no.~1,
  pp.~19--26, 2017.

\bibitem{Iwaszczuk.2012}
K.~Iwaszczuk {\em et~al.}, ``{Flexible metamaterial absorbers for stealth
  applications at terahertz frequencies},'' {\em {Optics Express}}, vol.~20,
  no.~1, p.~635, 2012.

\bibitem{PHASEDANTENNAS}
R.~J. Mailloux, {\em Phased array antenna handbook}, vol.~2.
\newblock Artech House Boston, 2005.

\bibitem{Yang.2016}
H.~Yang, X.~Cao, F.~Yang, J.~Gao, S.~Xu, M.~Li, X.~Chen, Y.~Zhao, Y.~Zheng, and
  S.~Li, ``A programmable metasurface with dynamic polarization, scattering and
  focusing control,'' {\em Scientific reports}, vol.~6, p.~35692, 2016.

\bibitem{PCBCART}
PCBcart, ``Printed circuit board calculator,'' 2017.

\bibitem{LAEBOOK}
M.~Caironi, {\em Large area and flexible electronics}.
\newblock John Wiley \& Sons, 2015.

\bibitem{LAEAPP}
M.~A.~U. Karim, S.~Chung, E.~Alon, and V.~Subramanian, ``Fully inkjet-printed
  stress-tolerant microelectromechanical reed relays for large-area
  electronics,'' {\em Advanced Electronic Materials}, vol.~2, no.~5, 2016.

\bibitem{LAEPRINTED}
R.~Parashkov, E.~Becker, T.~Riedl, H.-H. Johannes, and W.~Kowalsky, ``Large
  area electronics using printing methods,'' {\em Proceedings of the IEEE},
  vol.~93, no.~7, pp.~1321--1329, 2005.

\bibitem{LAEcost}
K.~Wiesenh\"{u}tter and W.~Skorupa, ``Low-cost and large-area electronics,
  roll-to-roll processing and beyond,'' in {\em Subsecond Annealing of Advanced
  Materials}, pp.~271--295, Springer International Publishing, 2014.

\bibitem{paper_print_cost}
``Printing cost calculator,'' May 2017.

\bibitem{Akyildiz.2008}
I.~F. Akyildiz {\em et~al.}, ``{Nanonetworks: A new communication paradigm},''
  {\em {Computer Networks}}, vol.~52, no.~12, pp.~2260--2279, 2008.

\bibitem{books2011automata}
J.~E. Hopcroft, R.~Motwani, and J.~D. Ullman, {\em Automata theory, languages,
  and computation}, vol.~24.
\newblock 2006.

\bibitem{Verikoukis.2017}
C.~Verikoukis {\em et~al.}, ``{Internet of Things: Part 2},'' {\em {IEEE
  Communications Magazine}}, vol.~55, no.~2, pp.~114--115, 2017.

\bibitem{Holloway.2012}
C.~L. Holloway {\em et~al.}, ``{An Overview of the Theory and Applications of
  Metasurfaces: The Two-Dimensional Equivalents of Metamaterials},'' {\em {IEEE
  Antennas and Propagation Magazine}}, vol.~54, no.~2, pp.~10--35, 2012.

\bibitem{nocarc18}
T.~Saeed {\em et~al.}, ``Fault adaptive routing in metasurface controller
  networks,'' in {\em Proceedings of the 11th International Workshop on Network
  on Chip Architectures (NoCArc'18)}, (Fukuoka, Japan), 2018.

\bibitem{haupt2007genetic}
R.~L. Haupt and D.~H. Werner, {\em {Genetic algorithms in electromagnetics}}.
\newblock John Wiley {\&} Sons, 2007.

\bibitem{nanocom.2017}
A.~Tsioliaridou, C.~Liaskos, A.~Pitsillides, and S.~Ioannidis, ``A novel
  protocol for network-controlled metasurfaces,'' in {\em ACM NANOCOM'17},
  NanoCom '17, (New York, NY, USA), pp.~3:1--3:6, ACM, 2017.

\bibitem{localization60ghzCmaccuracy}
J.~Chen {\em et~al.}, ``Pseudo lateration: Millimeter-wave localization using a
  single rf chain,'' in {\em Wireless Communications and Networking Conference
  (IEE WCNC)}, pp.~1--6, 2017.

\bibitem{ActixLtd.2010}
{Actix~Ltd}, ``{Radiowave Propagation Simulator SE},'' {\em http://actix.com},
  2010.

\bibitem{Yazdi.2017}
M.~Yazdi and M.~Albooyeh, ``{Analysis of Metasurfaces at Oblique Incidence},''
  {\em {IEEE Transactions on Antennas and Propagation}}, vol.~65, no.~5,
  pp.~2397--2404, 2017.

\bibitem{Albooyeh.2014}
M.~Albooyeh {\em et~al.}, ``Resonant metasurfaces at oblique incidence:
  interplay of order and disorder,'' {\em Scientific reports}, vol.~4, p.~4484,
  2014.

\bibitem{MatlabGA}
Mathworks, ``Genetic algorithm: Finding global optima for highly nonlinear
  problems,'' 2017.

\bibitem{Luke.2009}
S.~Luke, {\em {Essentials of metaheuristics}}.
\newblock [S.l.]: Lulu, 1~ed., 2009.

\bibitem{akyildiz2015internet}
I.~F. Akyildiz, M.~Pierobon, S.~Balasubramaniam, and Y.~Koucheryavy, ``{The
  internet of bio-nano things},'' {\em {IEEE Communications Magazine}},
  vol.~53, no.~3, pp.~32--40, 2015.

\bibitem{Fan.2013}
K.~Fan {\em et~al.}, ``{Optically Tunable Terahertz Metamaterials on Highly
  Flexible Substrates},'' {\em {IEEE Transactions on Terahertz Science and
  Technology}}, vol.~3, no.~6, pp.~702--708, 2013.

\end{thebibliography}

\parpic{\includegraphics[width=1in,clip,keepaspectratio]{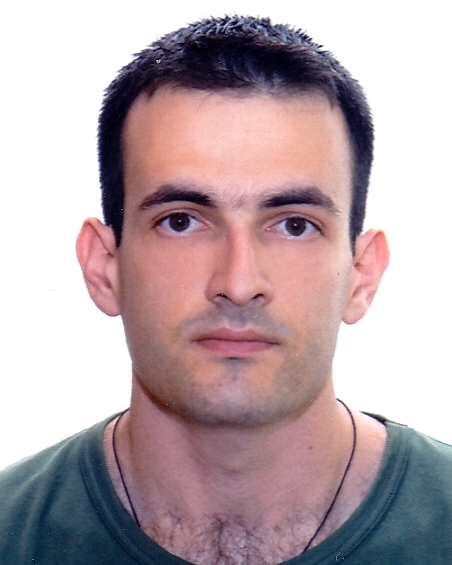}}
\noindent {\bf Christos Liaskos} received the Diploma in Electrical and Computer Engineering from the Aristotle University of Thessaloniki (AUTH), Greece in 2004, the MSc degree in Medical Informatics in 2008 from the Medical School, AUTH and the PhD degree in Computer Networking from the Dept. of Informatics, AUTH in 2014. He has published work in several venues, such as IEEE Transactions on: Networking, Computers, Vehicular Technology, Broadcasting, Systems Man and Cybernetics, Networks and Service Management, Communications, INFOCOM. He is currently a researcher at the Foundation of Research and Technology, Hellas (FORTH). His research interests include computer networks, security and nanotechnology, with a focus on developing nanonetwork architectures and communication protocols for future applications.

\parpic{\includegraphics[width=1in,clip,keepaspectratio]{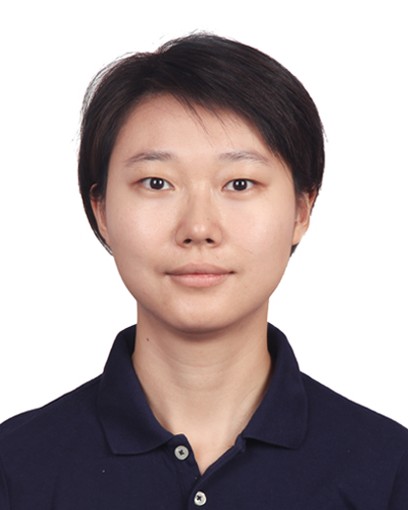}}
\noindent {\bf Shuai Nie} received the B.S. degree in Electrical Engineering from Xidian University in 2012, and the M.S. degree in Electrical Engineering from New York University in 2014. Currently, she is working toward the Ph.D. degree in electrical and computer engineering at the Georgia Institute of Technology under the supervision of Prof. Ian F. Akyildiz. Her research interests include terahertz band and millimeter-wave communication networks and the wireless communication system for 5G and beyond.

\parpic{\includegraphics[width=1in,clip,keepaspectratio]{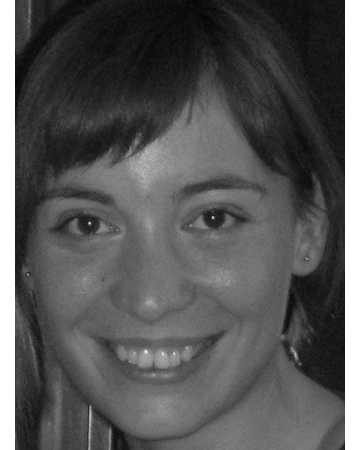}}
\noindent {\bf Ageliki Tsioliaridou} received the Diploma and PhD degrees in Electrical and Computer Engineering from the Democritus University of Thrace (DUTH), Greece, in 2004 and 2010, respectively. Her research work is mainly on the field of Quality of Service in computer networks. Additionally, her recent research interests lie in the area of nanonetworks, with specific focus on architecture, protocols,  security and authorization issues. She has contributed to a number of EU, ESA and National research projects. She is currently a researcher at the Foundation of Research and Technology, Hellas (FORTH).

\parpic{\includegraphics[width=1in,clip,keepaspectratio]{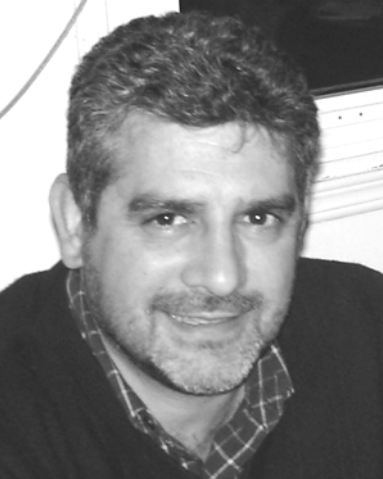}}
\noindent {\bf Andreas Pitsillides}  is a Professor in the Department of Computer Science, University of Cyprus, heads NetRL, the Networks Research Laboratory he founded in 2002, and is appointed Visiting Professor at the University of the Witwatersrand (Wits), School of Electrical and Information engineering, Johannesburg, South Africa. Earlier (2014-2017) Andreas was appointed Visiting Professor at the University of Johannesburg, Department of Electrical and Electronic Engineering Science, South Africa. His broad research interests include communication networks (fixed and mobile/wireless), Nanonetworks and Software Defined Metasurfaces/Metamaterials, the Internet- and Web- of Things, Smart Spaces (Home, Grid, City), and Internet technologies and their application in Mobile e-Services, especially e-health, and security. He has a particular interest in adapting tools from various fields of applied mathematics such as adaptive non-linear control theory, computational intelligence, game theory, and recently complex systems and nature inspired techniques, to solve problems in communication networks. Published over 270 referred papers in flagship journals (e.g. IEEE, Elsevier, IFAC, Springer), international conferences and book chapters, 2 books (one edited), participated in over 30 European Commission and locally funded research projects as principal or co-principal investigator, received several awards, including best paper, presented keynotes, invited lectures at major research organisations, short courses at international conferences and short courses to industry, and serves/served on several journal and conference executive committees.

\parpic{\includegraphics[width=1in,clip,keepaspectratio]{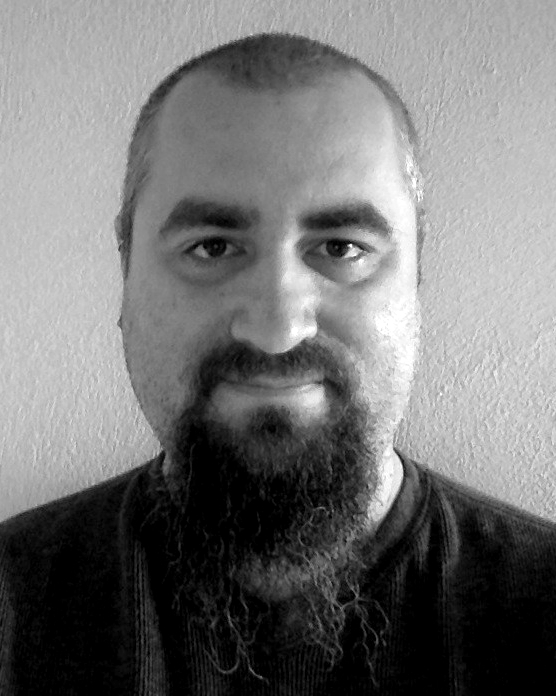}}
\noindent {\bf Sotiris Ioannidis} received a BSc degree in Mathematics and an MSc degree in Computer Science from the University of Crete in 1994 and 1996 respectively. In 1998 he received an MSc degree in Computer Science from the University of Rochester and in 2005 he received his PhD from the University of Pennsylvania. Ioannidis held a Research Scholar position at the Stevens Institute of Technology until 2007, and since then he is Research Director at the Institute of Computer Science of the Foundation for Research and Technology - Hellas. Since November 2017 he is a member of the European Union Agency for Network and Information Security (ENISA) Permanent Stakeholders Group (PSG). His research interests are in the area of systems, networks, and security. Ioannidis has authored more than 100 publications in international conferences and journals, as well as book chapters, including ACM CCS, ACM/IEEE ToN, USENIX ATC, NDSS, and has both chaired and served in numerous program committees in prestigious international conferences. Ioannidis is a Marie-Curie Fellow and has participated in numerous international and European projects. He has coordinated a number of European and National projects (e.g. PASS, EU-INCOOP, GANDALF, SHARCS) and is currently the project coordinator of the THREAT-ARREST, I-BiDaaS, BIO-PHOENIX, IDEAL-CITIES, CYBERSURE, and CERTCOOP European projects.

\parpic{\includegraphics[width=1in,clip,keepaspectratio]{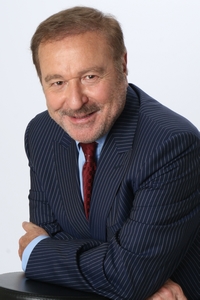}}
\noindent {\bf Ian F. Akyildiz} is currently the Ken Byers Chair Professor in Telecommunications with the School of Electrical and Computer Engineering, Director of the Broadband Wireless Networking Laboratory, and Chair of the Telecommunication Group at Georgia Institute of Technology, Atlanta, USA. Since 2011, he serves as a Consulting Chair Professor with the Department of Information Technology, King Abdulaziz University, Jeddah, Saudi Arabia, and with the Computer Science Department at the University of Cyprus since January 2017. He is a Megagrant Research Leader with the Institute for Information Transmission Problems at the Russian Academy of Sciences, in Moscow, Russia, since May 2018. His current research interests are in 5G wireless systems, nanonetworks, Terahertz band communications, and wireless sensor networks in challenged environments. He is an IEEE Fellow (1996) and an ACM Fellow (1997). He received numerous awards from the IEEE and the ACM, and many other organizations. His h-index is 115, and the total number of citations is above 105K as per Google scholar as of October 2018.

\end{document}